\documentclass{ieeeaccess}
\usepackage{cite}
\usepackage{amsmath,amssymb,amsfonts}
\usepackage{algorithmic}
\usepackage{algorithm}
\usepackage{graphicx}
\usepackage{textcomp}
\usepackage{bm}
\usepackage{autobreak}
\usepackage{wasysym}
\usepackage{array}
\usepackage{ulem}
\usepackage{url}
\usepackage{tabularx,booktabs}
\usepackage{flushend}
\def\vec#1{\mbox{\bf #1}}

\definecolor{myred}{rgb}{1,0.2,0.2}
\definecolor{myblue}{rgb}{0,0.3,1}
\definecolor{mygreen}{rgb}{0.2,0.7,0}
\definecolor{myorange}{rgb}{1,0.5,0}
\definecolor{mymagenta}{rgb}{1,0,1}

\def\BibTeX{{\rm B\kern-.05em{\sc i\kern-.025em b}\kern-.08em
    T\kern-.1667em\lower.7ex\hbox{E}\kern-.125emX}}

\begin{document}
\history{Date of publication xxxx 00, 0000, date of current version xxxx 00, 0000.}
\doi{10.1109/ACCESS.2023.1120000}

\title{Open Wireless Digital Twin: End-to-End 5G Mobility Emulation with OpenAirInterface and Ray Tracing}
\author{
\uppercase{Tetsuya Iye}\authorrefmark{1},
\uppercase{Masaya Sakamoto}\authorrefmark{2},
\uppercase{Shohei Takaya}\authorrefmark{1},
\uppercase{Eisaku Sato}\authorrefmark{1},
\uppercase{Yuki Susukida}\authorrefmark{1},
\uppercase{Yu Nagaoka}\authorrefmark{3},
\uppercase{Kazuki Maruta}\authorrefmark{4}, \IEEEmembership{Senior Member, IEEE}, and
\uppercase{Jin Nakazato}\authorrefmark{4}, \IEEEmembership{Member, IEEE}
}
\address[1]{Kozo Keikaku Engineering Inc., Tokyo, Japan}
\address[2]{Independent Researcher}
\address[3]{Department of Electronics and Computer Systems, Graduate School of Engineering, Takushoku University, Tokyo, Japan}
\address[4]{Department of Electrical Engineering, Graduate School of Engineering, Tokyo University of Science, Tokyo, Japan}
\tfootnote{This research and development work was supported in part by Fundamental Technologies for Sustainable Efficient Radio Wave Use R\&D Project (FORWARD) from the Ministry of Internal Affairs and Communication (receipt number JPMI240310001), Japan.}

\markboth
{T. Iye \headeretal: Open Wireless Digital Twin: End-to-End 5G Mobility Emulation with OpenAirInterface and Ray Tracing}
{T. Iye \headeretal: Open Wireless Digital Twin: End-to-End 5G Mobility Emulation with OpenAirInterface and Ray Tracing}

\corresp{Corresponding author: Tetsuya Iye (e-mail: tetsuya-iye@kke.co.jp).}

\begin{abstract}
This study presents an end-to-end wireless digital twin platform constructed using open-source software and open data to enhance the evaluation of mobile communication systems. The proposed open wireless digital twin (OWDT) integrates OpenAirInterface (OAI) for Fifth-Generation New Radio (5G NR) protocol stack emulation and NVIDIA Sionna RT for high-resolution ray-tracing-based radio propagation modeling. This integration enables the realistic emulation of 5G wireless communication in mobility scenarios on a CPU-based Linux system, leveraging real-world building data to bridge the gap between theoretical simulations and real-world deployments. The platform also incorporates OAI FlexRIC, which is an implementation aligned with the O-RAN near-real-time RAN Intelligent Controller (near-RT RIC), to dynamically monitor key performance indicators (KPIs). Through extensive evaluation in urban environments, this study demonstrated the validity of the emulation framework, revealing its capability to replicate real-world communication dynamics with high fidelity. The results underscore the potential of the OWDT to accelerate wireless system development, reduce experimental costs, and optimize network configurations.
\end{abstract}


\titlepgskip=-21pt

\maketitle

\section{Introduction}
\subsection{BACKGROUND}
\PARstart{I}{n} the research and development (R\&D) of wireless-assisted connected autonomous vehicles (CAVs) for future intelligent transportation systems (ITS), it is important to evaluate dynamic aspects of key performance indicators (KPIs) through demonstration experiments based on real-world usage scenarios~\cite{9690855,10637709}. Simultaneously, the demo constitutes one of the technical hurdles in wireless R\&D aimed at industrial applications. The challenges in such demo experiments include (1) discrepancies that arise between simulation models and real devices, (2) the need in some cases to obtain radio station licenses, which increases time and financial costs, (3) growing protocol complexity as systems become more advanced, and (4) the difficulty of accurately reproducing the radio-propagation environment during pre-evaluation of the demo scenarios.

To address challenges (1) to (4), an approach has been proposed in which scenario testing is conducted in a digital twin virtual environment that replicates the real world with high fidelity, enabling evaluation prior to field experiments~\cite{DigitalTwin,10365476}. Digital twins are being actively discussed in standardization projects, such as 3rd Generation Partnership Project (3GPP) and Open Radio Access Network (O-RAN). They not only enable the verification of new methods in virtual space prior to field demonstration experiments but also make it possible to reproduce time-varying KPIs that are difficult to observe in real communication systems and to optimize deployment parameters. Consequently, the development of digital twin platforms and their use for validation are attracting attention as promising methodologies for bridging the technical gaps among theory, simulation, and demonstration experiments, and for accelerating the cycle of translating research outcomes into societal deployment~\cite{WE1,WE2,WE3,colosseum,Boston,X5G,WDT,VDT,CCNC,ns3DT,ns3meets}. Furthermore, digital twins are regarded as a promising means of generating virtually unlimited training data for Artificial Intelligence / Machine Learning (AI/ML) models used for Radio Access Network (RAN) and network control~\cite{DT1,DT2,DTRAN}. Key technological elements for realizing digital twins include emulation technologies that guarantee high fidelity and real-time performance in virtual space, as well as consistency with measurement data that link the virtual and real worlds.

\begin{table}
\caption{\textbf{Comparison between current 5G/O-RAN OSS and this study}}
\label{table:oss}
\centering
\setlength{\tabcolsep}{4pt}
\begin{tabular}{wr{66pt}|wc{18pt}wc{18pt}wc{18pt}wc{18pt}wc{18pt}wc{18pt}}
\hline
Software Projects & CN & RAN & UE & OTA & Prop. & O-RAN \\
\hline
\hline
OAI\cite{oai,FlexRIC} & \checkmark & \checkmark & \checkmark & \checkmark & - & \checkmark\\
free5GC~\cite{free5GC} & \checkmark & - & - & - & - & - \\
Open5GS~\cite{open5GS} & \checkmark & - & - & - & - & - \\
Magma~\cite{Magma} & \checkmark & - & - & - & - & - \\
AETHER~\cite{AETHER} & \checkmark & \checkmark & - & - & - & \checkmark \\
srsRAN~\cite{srsRAN} & - & \checkmark & \checkmark & \checkmark & - & - \\
UERANSIM~\cite{UERANSIM} & - & \checkmark & \checkmark & - & - & - \\
Sionna~\cite{sionna,sionnaRT} & - & \checkmark & - & - & \checkmark & - \\
O-RAN SC~\cite{oransc} & - & \checkmark & - & - & - & \checkmark \\
5G-LENA~\cite{ns3LENA} & \checkmark & \checkmark & \checkmark & - & \checkmark & - \\
Simu5G~\cite{simu5G} & \checkmark & \checkmark & \checkmark & - & \checkmark & - \\
This Study & \checkmark & \checkmark & \checkmark & \checkmark & \checkmark & \checkmark \\
\hline
\multicolumn{7}{p{236pt}}{This compares only 5G OSS projects capable of independently implementing and experimenting with each 5G function without relying on external simulators. `OTA' denotes whether the framework can connect an RU and conduct OTA experiments; `Prop.' indicates the presence of stochastic or deterministic channel model implementations; `O-RAN' signifies that at least one component of the O-RAN architecture is implemented.}
\end{tabular}
\end{table}

In particular, regarding challenge (3), recent R\&D in wireless communications has shown significant advances in the development of various open-source software (OSS) and the use of open data. Driving this open innovation are the increasing sophistication, diversification, and complexity of wireless communication technologies; in fact, it has become difficult for a single organization to complete full-stack R\&D on its own. Many wireless technologies are standardized globally by standards bodies. Therefore, implementing these common specifications as OSS is an economically rational approach on a global scale.
From a user perspective, OSS allows the low-cost construction of R\&D environments while also enabling flexible customizations that are difficult to achieve with commercial software packages. As shown in Table~\ref{table:oss}, Fifth-Generation New Radio (5G NR) OSS implementations have emerged to perform functions related to the RAN, 5G Core Network (CN), User Equipment (UE), and O-RAN. Among them, OpenAirInterface (OAI) stands out as a widely used 5G OSS in R\&D, offering detailed implementations across 4G Long Term Evolution (LTE)/5G NR CN, RAN, UE, and O-RAN, along with support for over-the-air (OTA) experiments involving software-defined radio (SDR) devices and commercial off-the-shelf (COTS) O-RAN Radio Units (O-RUs)~\cite{oai,FlexRIC}. Free5GC, Open5GS, and Magma constitute three of the most prominent OSS implementations of the 5G CN~\cite{free5GC,open5GS,Magma}. AETHER, led by the Open Networking Foundation, is a project that unifies the subprojects SD-Core, SD-RAN, and an O-RAN near-real-time RAN Intelligent Controller (near-RT RIC) built on \textmu ONOS~\cite{AETHER}. The srsRAN project, including srsUE, delivers standalone RAN and UE functionality and enables OTA experimentation with SDRs or O-RUs~\cite{srsRAN}. UERANSIM is a software simulator that provides RAN and UE functionality for testing~\cite{UERANSIM}. NVIDIA Sionna is a GPU-accelerated link-level simulator for wireless communication systems~\cite{sionna}. In addition, Sionna RT offers differentiable ray tracing for deterministic radio propagation simulations~\cite{sionnaRT}. The O-RAN Software Community (O-RAN SC) supplies multiple components of the O-RAN architecture, including O-CUs, O-DUs, RICs, and an SMO framework~\cite{oransc}. 5G-LENA and Simu5G are discrete-event network simulators built on the ns-3 and OMNeT++ platforms, respectively~\cite{ns3LENA,simu5G}. A more detailed comparison of OSS projects from an O-RAN-focused perspective has been reported in a survey by another group~\cite{ORANOSS}.

The modeling of radio propagation, which falls under challenge (4), has been studied for improved accuracy using both stochastic and deterministic approaches. The stochastic approach allows the construction of channel models that statistically approximate experimental measurement results through parameter fitting. However, when 3D building models and terrain data are available, deterministic approaches, such as ray tracing, can explicitly track propagation paths and perform high-fidelity channel calculations from a microscopic perspective. In particular, ray tracing excels at simulating delay profiles in environments with significant multipath propagation. While multipath signals are treated as fading effects at the receiver, a trade-off exists between the number of delay taps of the signals and the computational processing cost. Real-time processing is essential for ensuring the emulation of digital twins. Therefore, efficient computation mechanisms are required to maintain the necessary number of delay taps to reproduce realistic fading effects while ensuring the real-time calculation of received signals.

\subsection{RELATED WORKS}
In the commercial sector, some companies have introduced network digital twin (NDT) solutions that incorporate proprietary AI technologies~\cite{Ericsson,VIAVI}. Here, `NDT' is defined as `virtual replica of mobile network or part of one, that captures its attributes, behavior and interactions,' noting that the mobile network includes both RAN and CN in 3GPP~\cite{DT1}.

A full-stack RAN simulation environment that integrates ns-3 and Sionna RT, considering deterministic channel models, was reported~\cite{ns3DT,ns3meets}. Although ns-3 is well-suited for scalable network performance simulation, it abstracts the physical (PHY) layer signal processing and lacks sub-microsecond discrete-time resolution. Therefore, there are inherent limits to capturing the time-domain effects of delayed multipath components, even when ray tracing is applied to model radio propagation.

Open AI Cellular (OAIC) is an NSF-supported OSS platform that stitches together srsRAN for RAN/UE operation and O-RAN SC components (near-RT/non-RT RIC and supporting services) to build an end-to-end (E2E), AI-ready 5G/6G research testbed~\cite{OAIC}. The OAIC supports the SDR operation and performs channel emulation via Zero Message Queue (ZeroMQ). In principle, if an efficient channel convolution is implemented for baseband In-phase and Quadrature (IQ) signal processing, it should be possible to emulate time-varying channels. However, the OAIC currently lacks 5G CN functionality and thus, the `E2E' emulation required in this study cannot be achieved.

Colosseum, a large-scale wireless network emulator developed at Northeastern University, implements an O-RAN-compliant digital twin environment~\cite{colosseum}. Its channel emulation is performed in the Massive Channel Emulator (MCHEM), a Field-Programmable Gate Array (FPGA) cluster that convolves channel impulse responses (CIRs) with baseband signals exchanged among numerous transmitter and receiver nodes. The convolution uses a 512‑tap finite impulse response (FIR) filter in which 4 arbitrary taps are set to nonzero values, a design that ensures high processing speed, lightweight data handling, and scalability with respect to the number of radio nodes. This 4‑tap approximation, however, limits the physical fidelity of each radio channel and forces the omission of bandwidth-limiting filtering that should, in principle, be applied to the CIR before convolution.

Another closely related effort has recently emerged within OAI~\cite{OAIemu}. Implemented on Godot, it enables channel generation via scenario control over a representational state transfer (REST) application programming interface (API) using the functions provided by Sionna~\cite{Godot}. With CIRs produced by this implementation, it is also possible to simulate 5G communications while accounting for dynamic channel updates via OAI \texttt{vrtsim}, a radio simulator with a fixed timescale. Our implementation differs in that it runs on a best-effort basis on the CPU in the simulation mode and operates only in real time in the SDR mode.

\subsection{CONTRIBUTIONS AND LIMITATIONS}
We summarize the contributions of the proposed Open Wireless Digital Twin (OWDT) architecture as follows. 
\begin{enumerate}
\item \textbf{An O-RAN-aligned 5G E2E evaluation environment.} We provide an evaluation environment positioned as shown in the bottom row of Table~\ref{table:oss} that integrates the strengths of OAI and Sionna RT to realize full-stack, non-abstracted 5G NR mobile communications in software. Here, `E2E' scope spans 5G CN, RAN, and UE. Furthermore, by leveraging OAI FlexRIC, KPIs can be monitored in near-RT~\cite{orancan}.
\item \textbf{A fully OSS-based wireless digital twin.} All software and data used to build the environment were open-source and open data. In addition, the CIR convolution implemented on OAI in this study and the sample scenarios were released as the OAI branch~\cite{OWDT}.
\item \textbf{High-fidelity wireless communication emulation.} By performing real-time baseband convolution with pre-computed CIRs using SDR, we achieved high-fidelity wireless emulation driven by site-specific mobility scenarios. Real-time convolution with up to 28 taps improves upon prior work~\cite{colosseum,WE2}.
\item \textbf{A low-cost, compact testbed.} As a testbed, the OWDT consists of a standard Linux server that runs on the CPU, a low-end SDR, and a COTS UE module, enabling the realization of a low-cost experimental platform.
\end{enumerate}
These contributions enable controlled yet realistic investigations of the PHY–propagation boundary within a laboratory setup, thereby promoting the democratization of wireless digital twin experimentation.

Meanwhile, the present OWDT has the following limitations, so the emulation remains approximate:
(a) It does not model channels varying on time scales shorter than the Doppler period;
(b) out of all the computed CIR taps, it uses only the dominant 28 taps. Although this is an expansion by several times over prior works, it remains a constraint;
(c) it has not yet been calibrated against measured data;
(d) prioritizing PHY fidelity comes at the expense of scalability, limiting the number of radio nodes that can be supported simultaneously.

\subsection{PAPER ORGANIZATION}
The remainder of this paper is organized as follows. Section~\ref{method} introduces the E2E system architecture of the OWDT, incorporating the 5G NR protocol stack implemented by OAI, FlexRIC enabling O-RAN-aligned key performance measurement (KPM), and propagation channels generated by Sionna RT. It also describes the hardware environment in which these components were deployed. Next, the workflow for executing mobility emulation scenarios, such as vehicular driving, on the OWDT platform is explained. This section also introduces the OSSs and open data utilized in the workflow. Furthermore, it provides a detailed discussion of the convolution method used to apply CIRs to baseband signals, which serves as the integration point between OAI and Sionna RT. Section~\ref{results} evaluates the data obtained through ray tracing and the KPIs measured from the OAI after applying these data to the system while executing the mobility scenarios on the platform. Finally, Section~\ref{conclusion} concludes the paper with a discussion of future perspectives. Table~\ref{tab:abbrev} summarizes the abbreviations used throughout the paper; all terms are defined at first use.
\begin{table*}[t]
\caption{\textbf{Abbreviations used in this paper.}}
\label{tab:abbrev}
\centering
\begin{tabular}{wl{45pt}wl{220pt}|wl{45pt}wl{150pt}}
\hline
\textbf{Acronym} & \textbf{Expanded form} & \textbf{Acronym} & \textbf{Expanded form} \\
\hline
3GPP & 3rd Generation Partnership Project & OAI & OpenAirInterface \\
5G NR & Fifth-Generation New Radio & OAIC & Open AI Cellular \\
5GCN & 5G Core Network & O\mbox{-}CU & O\mbox{-}RAN Central Unit \\
AGC & Automatic Gain Control & O\mbox{-}DU & O\mbox{-}RAN Distributed Unit \\
AI & Artificial Intelligence & O\mbox{-}RAN & Open Radio Access Network \\
AI/ML & Artificial Intelligence / Machine Learning & O\mbox{-}RAN SC & O\mbox{-}RAN Software Community \\
API & Application Programming Interface & O\mbox{-}RU & O\mbox{-}RAN Radio Unit \\
BLER & Block Error Rate & OFDM & Orthogonal Frequency Division Multiplexing \\
BLIS & BLAS-like Library Instantiation Software & OMNeT++ & Objective Modular Network Testbed in C++ \\
CAVs & Connected and Autonomous Vehicles & ONOS & Open Network Operating System \\
CDL & Clustered Delay Line & OSM & OpenStreetMap \\
CIR & Channel Impulse Response & OSS & Open-Source Software \\
CN & Core Network & OTA & Over-the-Air \\
COTS & Commercial Off-The-Shelf & OWDT & Open Wireless Digital Twin \\
CPU & Central Processing Unit & PDP & Power Delay Profile \\
CP & Cyclic Prefix & PHY & Physical (layer) \\
DL & Downlink & PLY & Polygon (file format) \\
E2 & E2 Interface & PRB & Physical Resource Block \\
E2E & End-to-End & R\&D & Research and Development \\
E2SM & E2 Service Model & RAN & Radio Access Network \\
FFT & Fast Fourier Transform & REST & Representational State Transfer \\
FIR & Finite Impulse Response & RF & Radio Frequency \\
FPGA & Field-Programmable Gate Array & RIC & RAN Intelligent Controller \\
FR2 & Frequency Range 2 & RMS & Root Mean Square \\
GI & Guard Interval & RSRP & Reference Signal Received Power \\
GPU & Graphics Processing Unit & RT & Ray Tracing \\
gNB & next-generation NodeB (5G base station) & SA & Standalone \\
IQ & In-phase and Quadrature & SDK & Software Development Kit \\
ISI & Inter-Symbol Interference & SDR & Software-Defined Radio \\
ITS & Intelligent Transportation Systems & SISO & Single-Input Single-Output \\
KPI & Key Performance Indicator & SRS & Sounding Reference Signal \\
KPM & Key Performance Measurement (O-RAN E2SM\mbox{-}KPM) & SUMO & Simulation of Urban MObility \\
LoS & Line-of-Sight & TDD & Time Division Duplex \\
LOD & Level of Detail & TDL & Tapped Delay Line \\
LTE & Long-Term Evolution & TR & Technical Report \\
LTS & Long-Term Support & TS & Technical Specification \\
MCHEM & Massive Channel Emulator & UE & User Equipment \\
MCS & Modulation and Coding Scheme & UL & Uplink \\
MIMO & Multiple-Input Multiple-Output & USRP & Universal Software Radio Peripheral \\
MLIT & Ministry of Land, Infrastructure, Transport and Tourism (Japan) & WSS & Wide-Sense Stationary \\
NDT & Network Digital Twin & xApp & O\mbox{-}RAN near-RT RIC Application \\
NR & New Radio & XML & Extensible Markup Language \\
ns\mbox{-}3 & network simulator 3 & ZeroMQ & Zero Message Queue \\
\hline
\end{tabular}
\end{table*}

\section{METHODS}\label{method}
This section outlines the architecture and operational workflow of the proposed platform. Subsection~\ref{ssct:arch} describes the modular architecture of the OWDT platform, detailing how OSSs and open data are integrated to create a realistic and flexible 5G emulation environment. Subsection~\ref{ssct:flow} explains the overall workflow of the OWDT platform, illustrating how data flow through the system from scenario setup and simulation to near-RT monitoring of KPIs.

\subsection{SYSTEM ARCHITECTURE}\label{ssct:arch}
We realized the OWDT by integrating the OAI, which reproduces the 5G NR protocol stack from the PHY layer and above, with NVIDIA Sionna RT v0.18.0, which enables the construction of a deterministic channel model through ray tracing. In this environment, CIRs pre-generated using Sionna RT were applied as propagation channels between the next-generation NodeB (gNB) and UE in the OAI. This allows us to simulate a scenario in which the UE moves within a 3D map while performing 5G communication in a laboratory setup. Furthermore, we can observe how KPIs, such as reference signal received powers (RSRPs), modulation and coding schemes (MCSs), block error rates (BLERs), and throughputs, dynamically change in the near-RT during the scenarios. These measurements were performed using the O-RAN standard KPM v3.00~\cite{KPM}, which is one of the service models supported by FlexRIC~\cite{FlexRIC}, a near-RT RIC that provides an xApp SDK framework developed within OAI.

\begin{figure}
    \centerline{\includegraphics[width=3.5in]{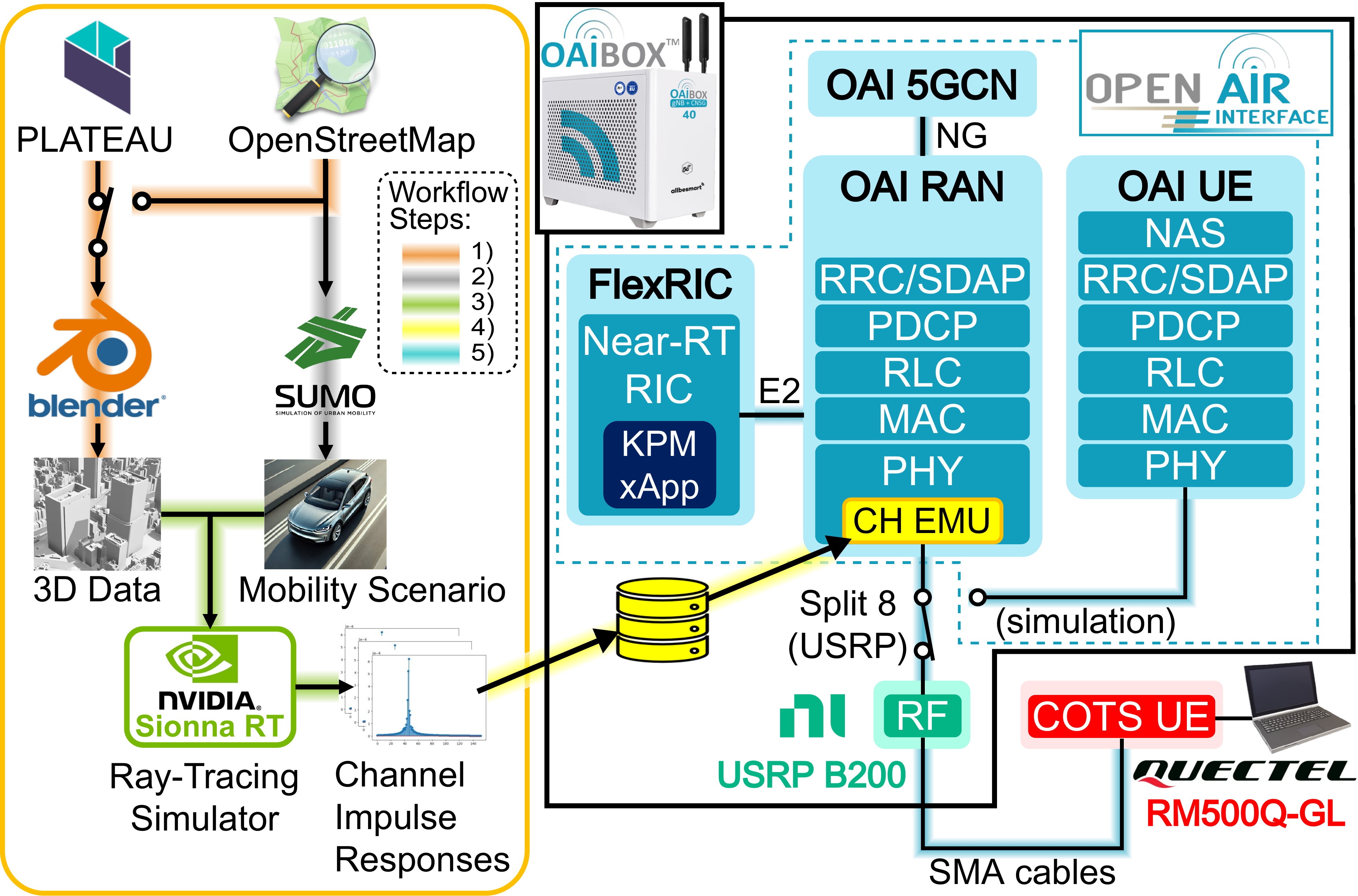}}
    \caption{Workflow of generating CIRs for mobility scenario (left panel) and 5G NR system architecture including 5GCN, RAN, Near-RT RIC, and UE developed by the OAI project, along with the RF devices serving as the RU and COTS UE (right panel).}
    \label{fig:architecture}
\end{figure}

The fundamental architecture of the OWDT is illustrated on the right side of Fig.~\ref{fig:architecture}. The hardware components that constitute the architecture are as follows: (i) A COTS server to implement the 5GCN, gNB, and near-RT RIC, along with an SDR device responsible for RF signal processing. The server was equipped with an AMD Ryzen 9 7900X 12-core processor (×24), 32 GiB RAM, and ran Ubuntu 22.04.4 LTS with Linux kernel version 6.8.1-060801-generic. An NI USRP B200 was used as the SDR device and connected via USB. In practice, this setup was realized using Allbesmart's OAIBOX\texttrademark\ 40~\cite{OAIBOX}. (ii) A computation server used for ray-tracing-based radio propagation simulation with Sionna RT. The server was equipped with an Intel(R) Xeon(R) Silver 4310 @ 2.10~GHz CPU, 256~GiB RAM, an NVIDIA A100 GPU (80~GB), and ran Ubuntu 22.04.4 LTS. (iii) A COTS laptop and SDR device pair for implementing an NR UE, or alternatively, any UE module that can be physically connected to the gNB via an RF cable. In this study, the Quectel RM500Q-GL UE module was connected to a laptop with an 11th Gen Intel Core i7-11370H @ 3.30~GHz (×8), 15.4~GiB RAM, and Ubuntu 20.04 LTS via USB.

\subsection{WORKFLOW}\label{ssct:flow}
An E2E wireless emulation following the mobility scenario was performed in the OWDT through the following five steps: 1) preparation of 3D building data, 2) generation of mobility scenarios, 3) channel impulse response calculation using ray tracing, 4) convolution of the channel impulse response, and 5) wireless communication emulation using OAI. In Fig.~\ref{fig:architecture}, the segments corresponding to each step are highlighted in different colors. First, an overview of each step is provided, followed by an explanation of the specific models and parameters adopted in this study.

\subsubsection{Preparation of 3D Building Data}
To achieve a higher resolution in radio propagation simulations, it is essential to prepare highly detailed 3D building data in terms of size and location and to perform site-specific deterministic evaluations. The 3D urban data provided by PLATEAU, an open data project promoted by the Ministry of Land, Infrastructure, Transport, and Tourism (MLIT) in Japan~\cite{plateau}, adopts building height measurements obtained via aerial surveying and generally provides higher accuracy in the height dimension than OpenStreetMap (OSM)~\cite{osm}. These data are available in the CityGML format, which cannot be directly handled by Sionna RT. Therefore, the open-source 3D computer graphics (CG) software Blender 4.2~\cite{blender} was used to convert the data into the PLY format. If OSM is used as 3D building data, it can be imported into Blender using Blosm~\cite{blosm}. The Mitsuba Blender Add-on~\cite{mitsuba} was used to assign materials to the polygon surfaces of the imported 3D model. In this study, two scenario areas were evaluated in Tokyo, Japan: Shin-nakano station and Shibuya station surroundings. A 3D model of the Shin-nakano station area is provided at Level of Details 1 (LOD1), where buildings are represented as solids, and all surfaces covering the solid are described as polygon elements. In contrast, the 3D model of the Shibuya station area is provided at LOD2, which allows for more detailed structures than LOD1, enabling the representation of walls, roofs, and floors with different materials. For the LOD1 Shin-nakano model, all materials, including roads, walls, and roofs, were designated as concrete, and the scene file was saved in XML format. In contrast, in the LOD2 Shibuya model, glass is assigned to building walls, concrete to road surfaces, and metal to overpass surfaces. These materials are determined using the values of the real part of the relative permittivity and conductivity, given by the equations $\epsilon_r = a (f\cdot10^{-9})^b$, $\sigma_c = c (f\cdot10^{-9})^d$, where $f$ is the carrier frequency, as described in~\cite{itur}. The specific parameter values for each material and the values of $\epsilon_r$ and $\sigma_c$ at the frequency used in this study are listed in Table~\ref{table:material}.
\begin{table}[t!]
\caption{Material properties for ray-tracing simulation~\cite{itur}}
\label{table:material}
\setlength{\tabcolsep}{3pt}
\centering
\begin{tabular}{wc{28pt}|wc{28pt}|wc{28pt}|wc{22pt}|wc{22pt}|wc{14pt}|wc{46pt}}
\hline
Material& \multicolumn{2}{c|}{Real part of} & \multicolumn{2}{c|}{Conductivity} & & \\ 
class   & \multicolumn{2}{c|}{relative permittivity} & \multicolumn{2}{c|}{[S/m]} & $\epsilon_{r}$ & $\sigma_c$ \\ \cline{2-5}
        & a & b & c & d & & (4.01916~GHz)\\
\hline
\hline
Vacuum & 1 & 0 & 0 & 0 & 1 & 0 \\
Concrete & 5.24 & 0 & 0.0462 & 0.7822 & 5.24 & 0.1372 \\
Glass & 6.31 & 0 & 0.0036 & 1.3394 & 6.31 & 0.0232 \\
Metal & 1 & 0 & $10^7$ & 0 & 1 & $10^7$ \\
\hline
\end{tabular}
\end{table}

\subsubsection{Generation of Mobility Scenarios}
The generation of mobility evaluation scenarios was handled by SUMO (Simulation of Urban MObility)~\cite{sumo}. SUMO is a vehicular traffic simulator capable of realistically replicating vehicle movement by incorporating traffic lights, multiple lanes, speed limits for each road, gradual acceleration and deceleration, and interactions with other vehicles. Based on OSM road data, SUMO enables the creation of vehicle mobility scenarios by specifying the driving lanes and their sequence and generating time-series updates of vehicle position coordinates $(x,y)$.
Although not used in this study, the mobility scenario can also include the velocity vector of the vehicle at each snapshot, which can be utilized to compute the Doppler shift as a key factor affecting fast fading. In both the Shin-nakano and Shibuya scenarios, vehicles temporarily stopped at traffic lights at intersections and made left turns. Additionally, in the narrow residential streets of Shin-nakano, where intersections and T-junctions lack traffic lights, vehicles either slow down or make a temporary stop before proceeding straight or making a left turn.

\subsubsection{Channel Impulse Response Calculation Using Ray Tracing}
Using any ray-tracing simulator, a deterministic and site-specific channel model for mobility can be computed by integrating the mobility scenario generated by SUMO with 3D urban geometry data from the OSM or PLATEAU. At arbitrary time intervals, the CIR can be calculated by considering reflections, diffractions, and diffuse scatterings between the gNB, which is fixed at a specific location, and the UE, which moves according to the mobility scenario.

NVIDIA Sionna RT (v0.18.0) is an OSS that leverages GPU acceleration based on TensorFlow to perform fast ray-tracing simulations of propagation environments~\cite{sionnaRT}. Sionna RT supports two channel models defined in 3GPP TR 38.901: the Clustered Delay Line (CDL) model and the Tapped Delay Line (TDL) model~\cite{TDLCDL}. Both express multipath propagation as a ``tapped delay line,'' but they differ in that the CDL treats each tap as a cluster and retains angular information, whereas the TDL considers only the delay and power of the taps in the time domain. We used a single isotropic antenna at both the gNB and the UE and did not evaluate Multiple-Input Multiple-Output (MIMO) or beamforming. Consequently, we did not use ray/cluster angles of arrival or antenna/beam directivity, and instead adopted a simpler TDL model, while also avoiding the increased processing load associated with multi-stream transmission. Sionna RT was used to efficiently conduct ray tracing over hundreds of snapshots throughout the scenario. However, other ray-tracing simulators that can output CIR, such as Wireless InSite and Opal, can also be applied to the same scenario~\cite{WI,opal,winprop,mwrt,withray}. A performance comparison across specific versions of several major ray-tracing simulators, including Sionna RT, was reported in a previous study~\cite{sionnaWIMW}. The resolution of the CIR output from the simulation depends on the 3D model and mobility scenarios. In particular, in complex urban environments, achieving a high-fidelity digital twin is feasible only with high-quality spatiotemporal data sets.

\begin{table}
\caption{System parameters for Sionna RT and OAI.}
\label{table:param}
\setlength{\tabcolsep}{3pt}
\centering
\begin{tabular}{wl{35pt}wl{105pt}wl{80pt}}
\hline
Simulation & Parameter & Value (Units) \\
\hline
\hline
Common & Frequency band & n77 \\
 & Carrier frequency $f$ & 4.01916~GHz \\
 & Number of transmitters $M_\text{T}$ & 1 \\
 & Number of receivers $M_\text{R}$ & 1 \\
 & Number of transmit antenna $N_\text{T}$ & 1 \\
 & Number of receive antenna $N_\text{R}$ & 1 \\
 & Sampling rate $f_\text{samp}$ & 46.08~Msps \\
\hline
Sionna RT & Antenna pattern $g(\theta,\phi)$ & 1 (0~dBi: isotropic) \\
 & Channel model & TDL \\
 & UE antenna height & 1.5~m \\
 & gNB antenna height   & 22.0~m (Shin-nakano) \\
 &                      & 10.5~m (Shibuya) \\
 & Reflection & True \\
 & Diffuse scattering & False\\
 & Max depth & 5 \\
 & Diffraction & True \\
 & Number of rays & $10^7$ \\
 & Coverage map resolution & 1 m $\times$ 1 m \\
 & Coverage map height & 1.5 m \\
 & Scenario interval $t_\text{int}$ & 100~ms \\
 & Number of snapshots & 570 (Shin-nakano) \\
 &                     & 230 (Shibuya) \\
\hline
OAI & Technology & 5G NR SA \\
 & Channel bandwidth $BW$ & 40~MHz \\
 & Subcarrier spacing & 30~kHz \\
 & FFT size & 1536 \\
 & Antenna & SISO \\
 & Duplex mode & TDD \\
 & TDD slot pattern & DDDSU \\
 & Special slot format & 6D+4G+4U \\
\hline
\end{tabular}
\end{table}
The simulation parameters for ray tracing are listed in Table~\ref{table:param}. The common parameters for both radio propagation and system models include the frequency band, center frequency, number of transceivers, number of antennas per transceiver, and sampling rate. In this study, we adopt n77 band (3.3~GHz--4.2~GHz), with a center frequency of 4.01916~GHz. The number of transceivers and antennas per transceiver were set to one. The sampling rate was set to $f_\text{samp}=46.08$~Msps, as supported by the USRP B200 used on the gNB. This sampling rate is also provided as an input argument to the Sionna RT functions when a bandwidth limitation is applied to the delay profile.

For the ray-tracing simulation, an isotropic antenna pattern was assumed, with $g(\theta, \phi)=1$, that is, 0~dBi for all directions. Reflection is enabled by setting \texttt{True}, and the maximum number of reflections is set to five by defining $\texttt{max\_depth}=5$. When diffraction is set to \texttt{True}, the model considers up to one diffraction per path without any reflection. In other words, the combination of reflections and diffractions was not calculated. 

\begin{figure}[t]
\centerline{\includegraphics[width=3.5in]{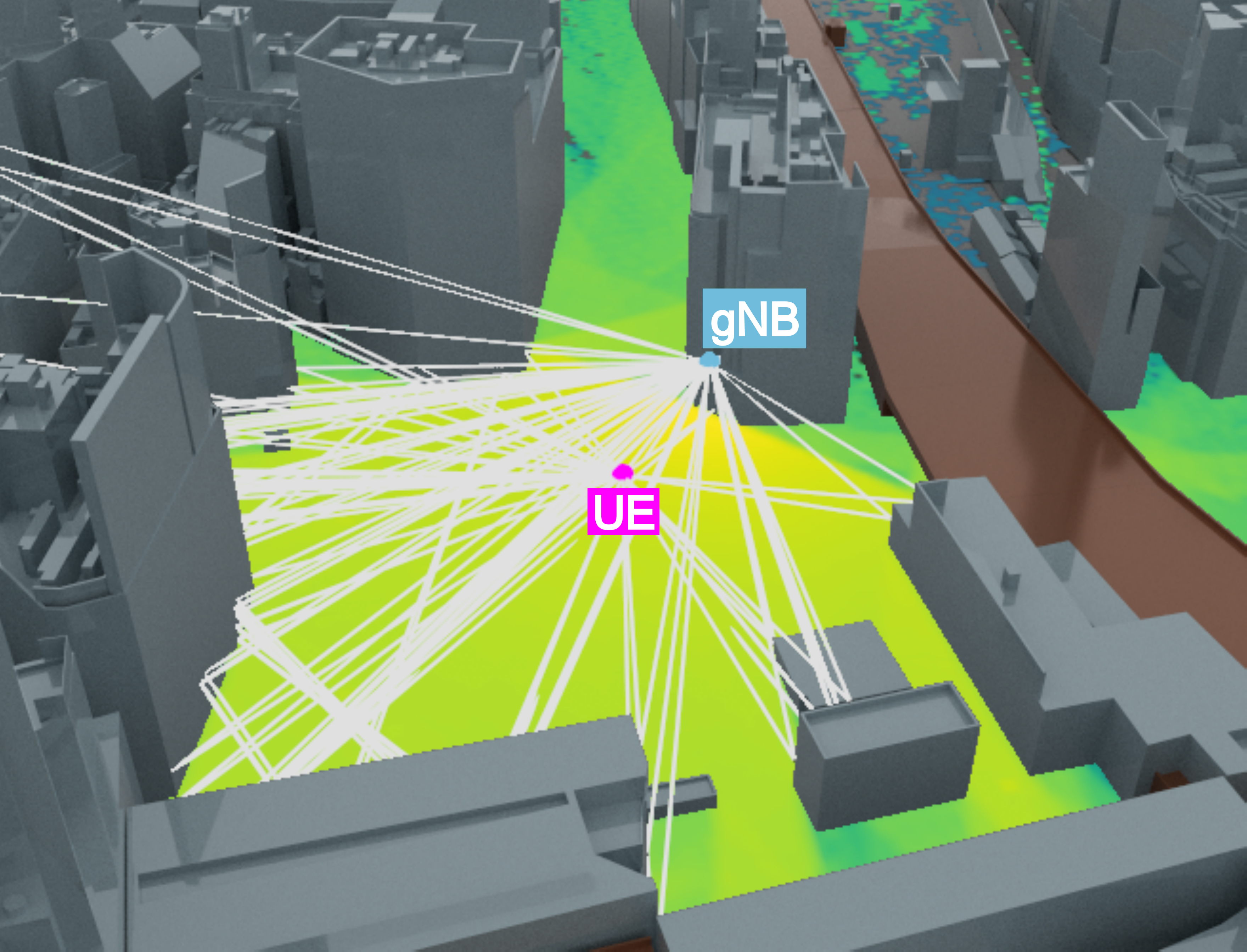}}
\caption{Ray-tracing simulation for Shibuya scenario.\label{fig:ray}}
\end{figure}

For ray tracing, the shoot-and-bounce approach with the ``fibonacci'' path search method~\cite{sionnaRT} was used, setting the number of candidate traced rays to $10^7$. The complex amplitude and delay of the gNB–UE link were computed for each snapshot by updating the UE coordinates every 100~ms according to the mobility scenario. For the Shin-nakano scenario, 570 snapshots were obtained over 57.0~s, while the Shibuya scenario involved 230 snapshots over 23.0~s. The total computation time for ray tracing across all snapshots was 13.5~s for Shin-nakano and 8.9~s for Shibuya. A snapshot of the Shibuya scenario is shown in Fig.~\ref{fig:ray}.

A radio propagation channel model is used to generate a delay profile, that is, complex amplitude $a_{p}\in\mathbb{C}$ and delay $\tau_{p}\in\mathbb{R}$ for each link between the transmitter and receiver antennas for $p^\text{th}$-path among $P\in\mathbb{N}$ paths. The amplitudes and delays were assumed to be time-independent. Such a delay profile against the delay time $\tau$ corresponds to be
\begin{equation}
    h(\tau)=\sum_{p=0}^{P-1}a_{p}\delta(\tau-\tau_{p}),
\end{equation}
where $\delta(\cdot)$ is the Dirac delta measure. The resulting discrete-time CIR under bandwidth limitation, assuming a sinc filter for pulse shaping and receiving filtering, is computed as follows:
\begin{equation}
\label{eq:barh}
    \bar{h}[k]=\sum_{p=0}^{P-1}a_{p} \ \text{sinc}(k-f_\text{samp}\tau_{p}),
\end{equation}
where $k\in\mathbb{Z}$ is the delay index corresponding to the sampling interval $1/f_\text{samp}$. Because the ideal sinc filter violates causality and is non-time-limited, it is necessary to consider from $k=-\infty$ to $k=+\infty$. However, in practice, these lower and upper limits need to be set to reasonable finite values, which can be computed using the Sionna RT \texttt{time\_lag\_discrete\_time\_channel()} function from a given sampling rate and a sufficiently reasonable maximum delay spread $3$ \textmu s to be $k=-6$ and $k=L_\text{max}-1=145$. For simplicity, we set the lower limit of the considered delay index to $k=0$.

\subsubsection{Convolution of Channel Impulse Response}
In the OAI software, wireless emulation can be executed by convolving the pre-computed CIR with the baseband signal in real time according to the time progression of the mobility scenario, thereby reflecting the propagation environment in the transmitted and received signals. This process allows the time evolution of the channel, owing to changes in the position of the moving object, to be reflected in the KPIs obtainable from the OAI.

\begin{figure}
\centerline{\includegraphics[width=3.5in]{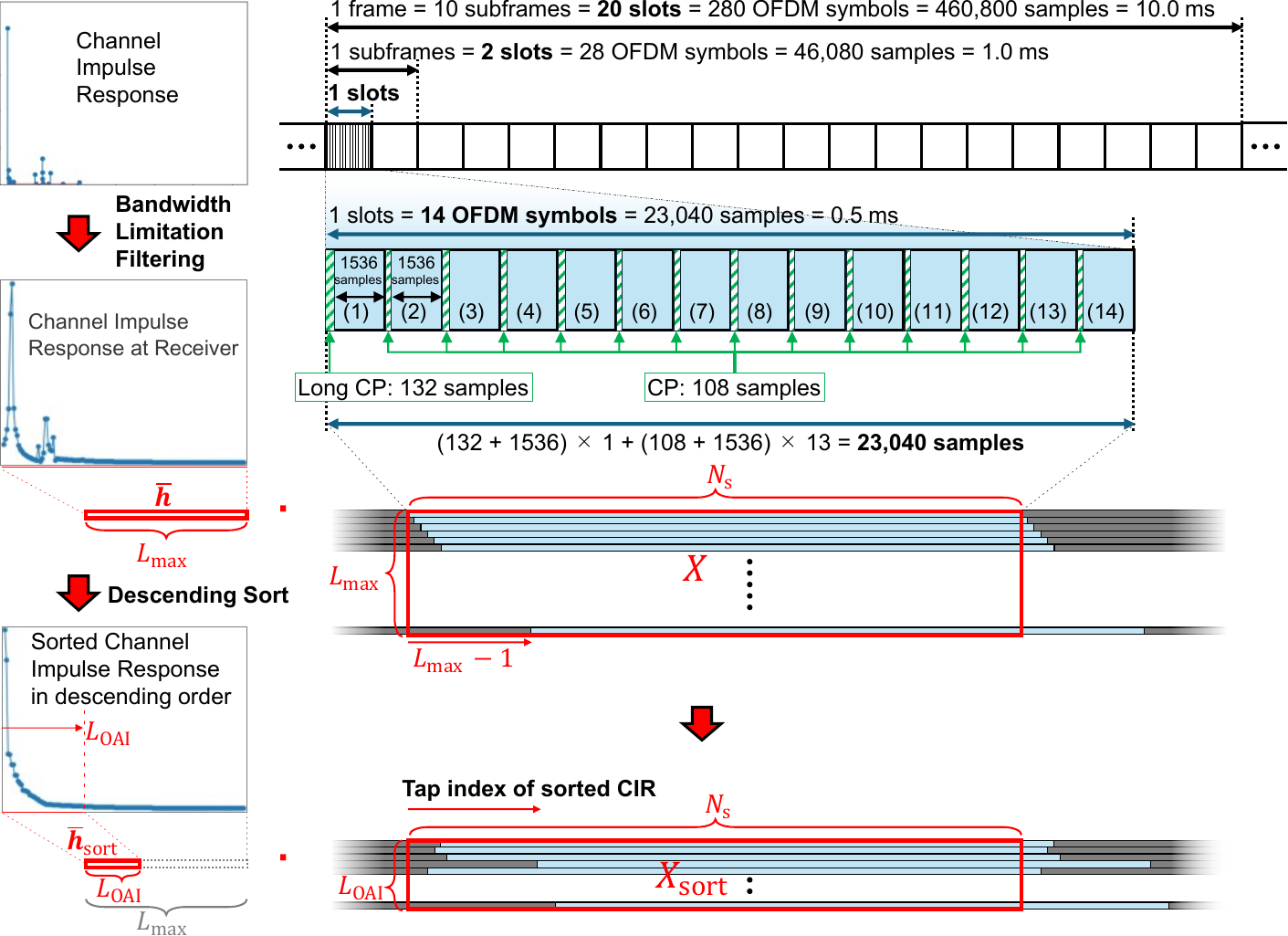}}
\caption{Implementation of CIR convolution for the proposed OWDT in OAI. 
\label{fig:CIRcalc}}
\end{figure}
In the OAI source code, baseband IQ samples corresponding to the lowest part of the PHY layer are handled within the \texttt{tx\_rf()} and \texttt{rx\_rf()} functions in the \texttt{nr-ru.c} file. The 5G NR frame format at the baseband IQ sample level and the CIR convolution method are shown in Fig.~\ref{fig:CIRcalc}. In the OAI RAN, the time-domain signal is stored in every slot as \texttt{txData} and \texttt{rxData}, where the baseband signal of Cyclic Prefix Orthogonal Frequency Division Multiplexing (CP-OFDM) is represented as $N_s(=\text{FFT size}\times15)$ complex IQ sample data. To incorporate the delay tap effects of a CIR snapshot into the baseband signal, a matrix $X\in\mathbb{C}^{L_\text{max}\times N_s}$ is generated by shifting the baseband signal sample-by-sample for the total number of CIR taps $L_\text{max}(=146)$. Applying the complex CIR vector $\bar{\vec{h}}\in\mathbb{C}^{L_\text{max}}$ from the left and computing the inner product results in the received signal $\vec{y}\in\mathbb{C}^{N_s}$ with multipath delay components. However, in our environment, even with the use of the high-performance computation library `BLIS' (described later), real-time convolution of the $L_\text{max}$-tap delay components obtained from ray tracing was not feasible. 

Therefore, considering the most significant power first, the CIR $\bar{\vec{h}}$ is sorted in descending order as $\bar{\vec{h}}_\text{sort}\in\mathbb{C}^{L_\text{OAI}}$, and the rows of the baseband delay matrix $X$ are reordered accordingly to form $X_\text{sort}\in\mathbb{C}^{L_\text{OAI}\times N_s}$. Here, the number of leading taps to be read is parameterized as $L_\text{OAI}$ to determine the upper limit of the real-time executable. This approach allows the computation to reflect the most significant multipath components first. The convolution operation was implemented within the \texttt{rx\_rf()} and \texttt{tx\_rf()} functions on the OAI gNB using \texttt{cblas\_cgemm} from the high-performance matrix computation library `BLIS'~\cite{blis}. Specifically, convolution was applied to the baseband signal before cable transmission from the gNB for the downlink (DL) and after cable reception from the UE for the uplink (UL). The function \texttt{cblas\_cgemm} multiplies two complex matrices $AB$ and multiplies the resulting matrix by a complex constant $\alpha$. Matrix $C$ is multiplied by another complex constant $\beta$. It stores the sum of these two products in matrix $C$ as follows: $C\leftarrow \alpha AB + \beta C$. We used this function to perform the following calculations.
\begin{equation}
\label{eq:y}
    \vec{y}=s\bar{\vec{h}}X+\sigma\vec{w}\approx s\bar{\vec{h}}_\text{sort}X_\text{sort}+\sigma\vec{w},
\end{equation}
where $s\in\mathbb{R}$ is the signal amplitude gain, $\sigma\in\mathbb{R}$ is the noise amplitude gain, and $\vec{w}\in\mathbb{C}^{N_s}$ is normalized complex Gaussian noise. To examine this equation in more detail, we express it at the sample time level as follows:
\begin{equation}
\label{eq:yk}
    y[n]=\sum_{k=0}^{L_\text{OAI}-1}s\bar{h}[k]x_k[n-k]+\sigma w[n],
\end{equation}
where $n\in\mathbb{Z}$ denotes the sampling time index. 

In general, in a time-invariant multipath delay environment, channel reciprocity is valid owing to the absence of Doppler spread~\cite{Karasawa}. Thus, in this mobility scenario, the theoretical validity of overlaying the same CIR on both the DL and UL in the OAI gNB is ensured.

To reproduce the effects of the CIR on the transmitted and received signals more accurately, it is necessary to perform convolution with a time-varying CIR for each IQ sample, considering the Doppler effect. However, from a computational cost perspective, real-time execution is infeasible without hardware acceleration, such as FPGA or GPU~\cite{RTCE}. In this study, we update the CIR every 100 ms and achieve real-time, E2E wireless emulation using only a commercial CPU-based server and a commercial SDR without any accelerators. As noted earlier, because OAI’s signal processing operates on a slot basis, the convolution with the CIR is executed once every 0.5 ms with the current parameter settings. In other words, the same CIR data is reused for 200 consecutive slots.

\subsubsection{Wireless Communication Emulation Using OAI}
The system parameters configured in the OAI are summarized in the lower part of Table~\ref{table:param}. The bandwidth $BW$ was set to 40~MHz, which is the maximum achievable bandwidth within the constraints of the sampling rate. The subcarrier spacing is set to 30~kHz, corresponding to numerology $\mu=1$. The 1,536‑point FFT employed here uses a three-quarter sampling frequency of the conventional $2^{n}$-sample base clock, thereby avoiding unnecessary bandwidth consumption; this energy-efficient FFT option is described in 3GPP TR 25.814~\cite{3/4sample}. OAI supports this FFT length when operating a 40~MHz channel on a USRP B200 configured for a 46.08~Msps sampling rate. Since the focus of this paper is first to demonstrate that the wireless communication system and the propagation phenomena can be simulated/emulated in a synchronized manner, we simplified the problem formulation as much as possible; the system operates in a single-input single-output (SISO) configuration. The duplex mode is Time Division Duplex (TDD), and the slot pattern follows `DDDSU.’ The special slot format follows a typical pattern with 6 DL symbols, 4 guard symbols, and 4 UL symbols.

\section{RESULTS AND DISCUSSION}\label{results}
\subsection{RAY-TRACING SIMULATION RESULTS}
\begin{figure}
\centerline{\includegraphics[width=3.5in]{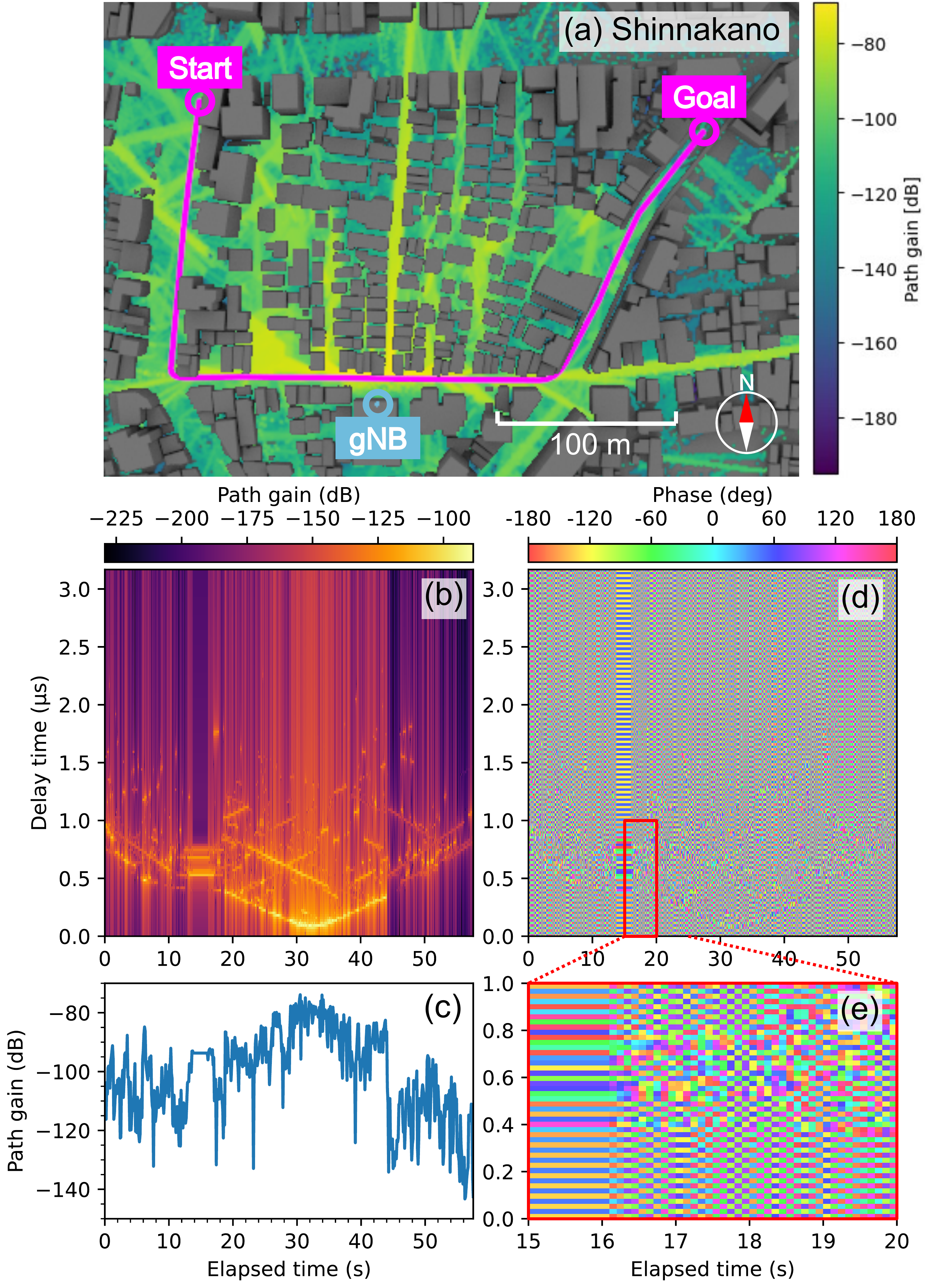}}
\caption{(a) Vehicle route on path gain coverage map for Shin-nakano scenario, (b) Power delay profile, (c) path gain, (d) phase delay profile, and (e) enlarged view of (d) for Shin-nakano scenario. \label{fig:shinnakano}}
\end{figure}

\begin{figure}
\centerline{\includegraphics[width=3.5in]{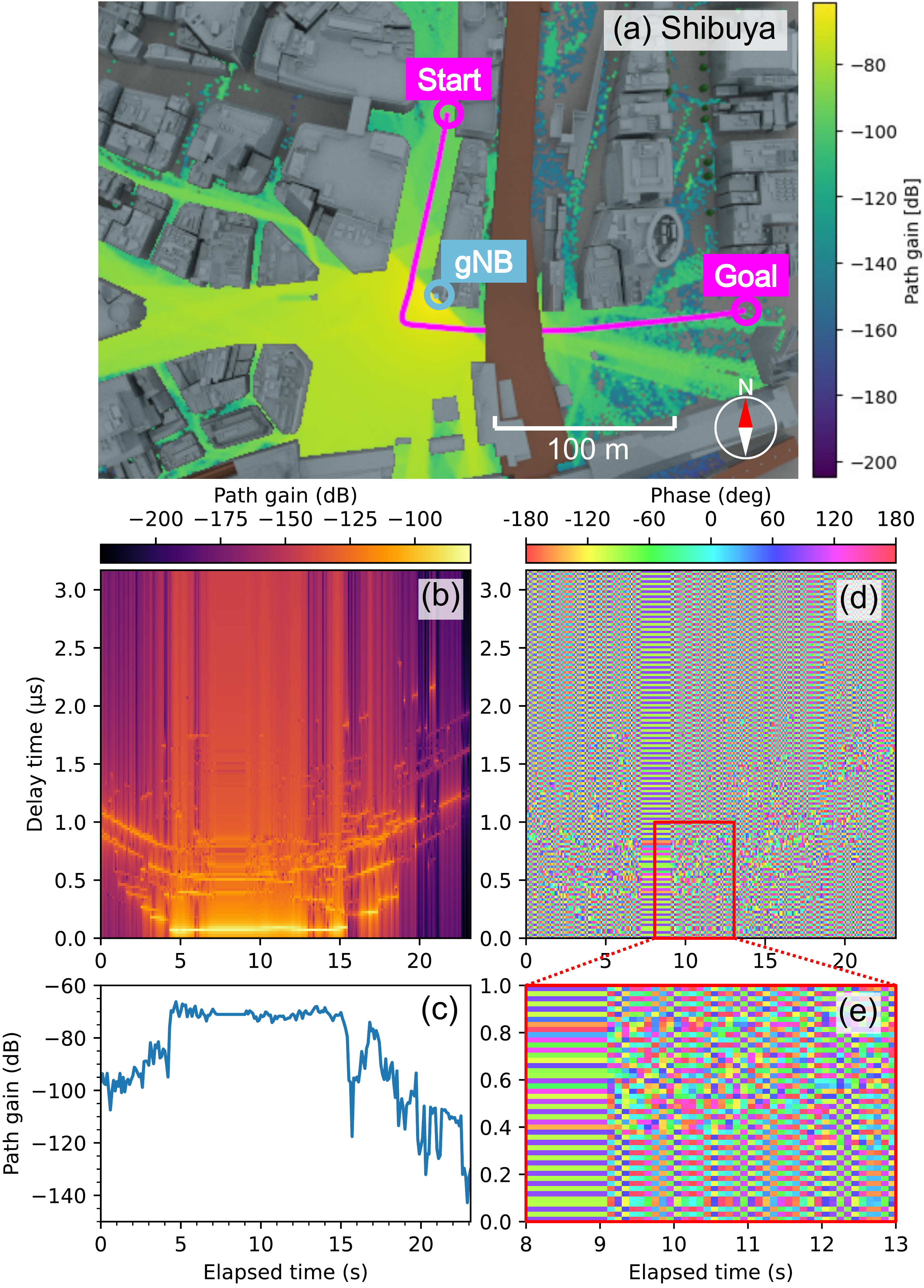}}
\caption{(a) Vehicle route on path gain coverage map for Shibuya scenario, (b) Power delay profile, (c) path gain, (d) phase delay profile, and (e) enlarged view of (d) for Shibuya scenario. \label{fig:shibuya}}
\end{figure}

Fig.~\ref{fig:shinnakano}(a) and Fig.~\ref{fig:shibuya}(a) show bird-eye views of the 3D models of Shin-nakano and Shibuya, depicting the fixed gNB position and vehicle movement trajectory overlaid with a path gain heatmap, respectively. The heatmaps are drawn on the horizontal planes at $z =$ 1.5 [m], matching the height of the moving UE. The paths considered in the path gain coverage map and CIR calculation include the line-of-sight (LoS) path, diffraction paths with up to one diffraction event, and specular reflections up to the number of reflections specified by \texttt{max\_depth}.

In the Shin-nakano scenario, the radio propagation environment was complex owing to the presence of residential, low-rise buildings, and narrow streets. Additionally, the 3D model provided by PLATEAU for the Shin-nakano area is at LOD1, resulting in relatively simple building shapes. In this mobility scenario, the vehicle started moving from a location far from the gNB at $t = 0$ and stopped briefly at a traffic light between $t = 13.8$ and $t = 16.0$. Upon receiving a green light, it makes a left turn and enters the street in front of the building where the gNB is installed. The vehicle continues eastward, passing in front of the gNB, briefly stops at another intersection, and then makes another left turn. It then proceeds to a shopping street heading north-northeast. 
This scenario consisted of 570 snapshots, corresponding to 57.0~s of mobility data. Using the CIR obtained every 100 ms, the power delay profile (PDP), calculated as the square of the complex amplitude, is shown in Fig.~\ref{fig:shinnakano}(b), and the phase delay profile is shown in Fig.~\ref{fig:shinnakano}(d). At approximately $t = 18$ [s], as the vehicle enters the street in front of the gNB, the LoS path component appears. As the vehicle approaches the gNB, the signal strength increases, and the delay decreases. When the vehicle passes the gNB and moves away, the signal strength weakens, the delay increases, and the LoS component exhibits a V-shaped pattern on the PDP. After the second left turn, not only does the LoS component diminish, but the number of reflection paths also drops significantly, resulting in approximately 30~dB of path gain attenuation. The change in the total path gain, obtained by summing all the complex amplitudes in the CIR, is shown in Fig.~\ref{fig:shinnakano}(c). The results clearly show a plateau period during the stop, a maximum value at approximately $t = 32$ [s], and a sharp 30~dB drop at approximately $t = 44$ [s].
Next, we discuss the phase behavior shown in Figs.~\ref{fig:shinnakano}(d) and (e). Fig.~\ref{fig:shinnakano}(d) indicates rapid phase fluctuations for each tap along the vertical delay time axis. As shown in the magnified view in Fig.~\ref{fig:shinnakano}(e), the phase remained constant during the stop and then started fluctuating once the vehicle began moving. The phase exhibited a consistent trend along the LoS path trajectory, as observed in the PDP. However, it should be noted that this study does not account for time variations within the 100~ms intervals; consequently, the phase values show discontinuous transitions in each snapshot. This indicates that our study did not consider the Doppler effect.

The Shibuya scenario shown in Fig.~\ref{fig:shibuya}(a) involves a route where the vehicle turns left in front of the gNB antenna installed at the corner of an intersection, passes under a brown-colored overpass, and moves away from the gNB. The differences from the Shin-nakano scenario include wider roads and building facades set as glass surfaces, representing an urban business district. In the mobility scenario, the vehicle started moving from the north and stopped briefly at the scrambled intersection because of a traffic signal between $t = 7.2$ [s] and $t = 9.1$ [s]. When the light turned green, the vehicle began a left turn and continued eastward in a straight line. The PDP obtained from the CIR is illustrated in Fig.~\ref{fig:shibuya}(b), and the phase delay profile is shown in Fig.~\ref{fig:shibuya}(d). At $t = 4.2$ [s], as the vehicle enters the gNB’s LoS range, the path gain sharply increases and maintains a plateau until approximately $t = 15.4$ [s], when the LoS component disappears owing to the building shadow, causing a sudden drop in path gain. The bright stepped lines visible in the PDP before and after the LoS interval represent first- and second-order reflections, whereas higher-order reflections are visible only before the plateau. By examining the magnified view in Fig.~\ref{fig:shibuya}(e), similar to the Shin-nakano scenario, the phase remains stable during the stop and begins to fluctuate as the vehicle starts to move.

\subsection{OAI KPM RESULTS}
By installing and running FlexRIC, which functions as an O-RAN-aligned near-RT RIC, various KPIs can be obtained in compliance with the E2 Service Model (E2SM) KPM v03.00. In this study, we focus on the RSRP, MCS, BLER, and throughput and discuss their behavior in response to the mobility scenario.

Before applying the time variations of the channel induced by the mobility scenario, we performed a baseline measurement as a reference using the same experimental setup, as shown in the right side of Fig. \ref{fig:architecture} without applying any CIR. This means that, in the baseline measurement, $|s|=|\bar{h}|=1$ in (\ref{eq:yk}). The 100-second average values of the KPIs are summarized in Table \ref{table:baseline}.
\begin{figure}
\centerline{\includegraphics[width=3.5in]{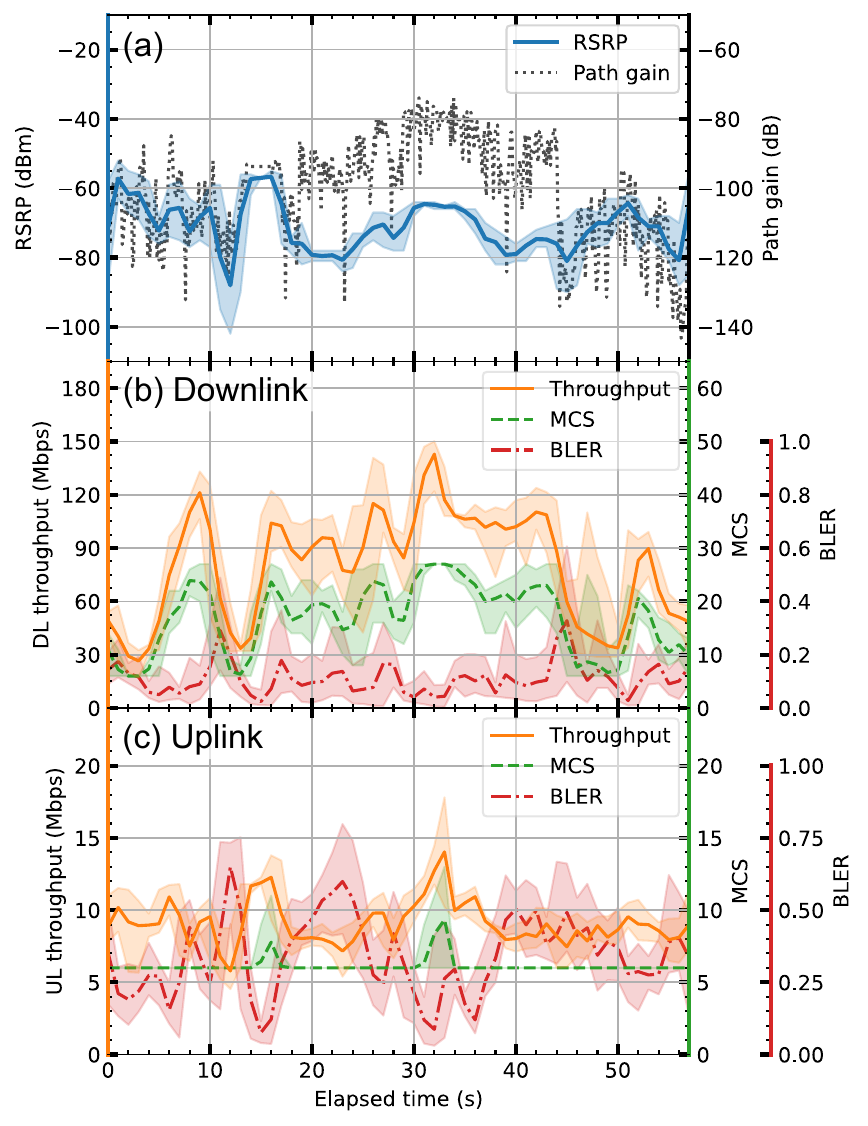}}
\caption{(a) RSRP with path gain obtained from RT simulation as a reference, (b) DL KPIs, and (c) UL KPIs obtained from OAI for Shin-nakano scenario.\label{fig:shinnakanoOAI}}
\end{figure}
\begin{table}
\caption{The 100-second averaged results of the baseline KPI measurements for DL and UL.}
\label{table:baseline}
\setlength{\tabcolsep}{3pt}
\centering
\begin{tabular}{wl{70pt}wl{60pt}}
\hline
KPI & Value (Units) \\
\hline
\hline
RSRP & -63~dBm \\
DL MCS & 27 \\
DL BLER & $2.62\cdot 10^{-5}$ \\
DL Throughput & 150.14~Mbps \\
UL MCS & 15.21 \\
UL BLER & 0.0963 \\
UL Throughput & 26.29~Mbps \\
\hline
\end{tabular}
\end{table}

\begin{figure}
\centerline{\includegraphics[width=3.5in]{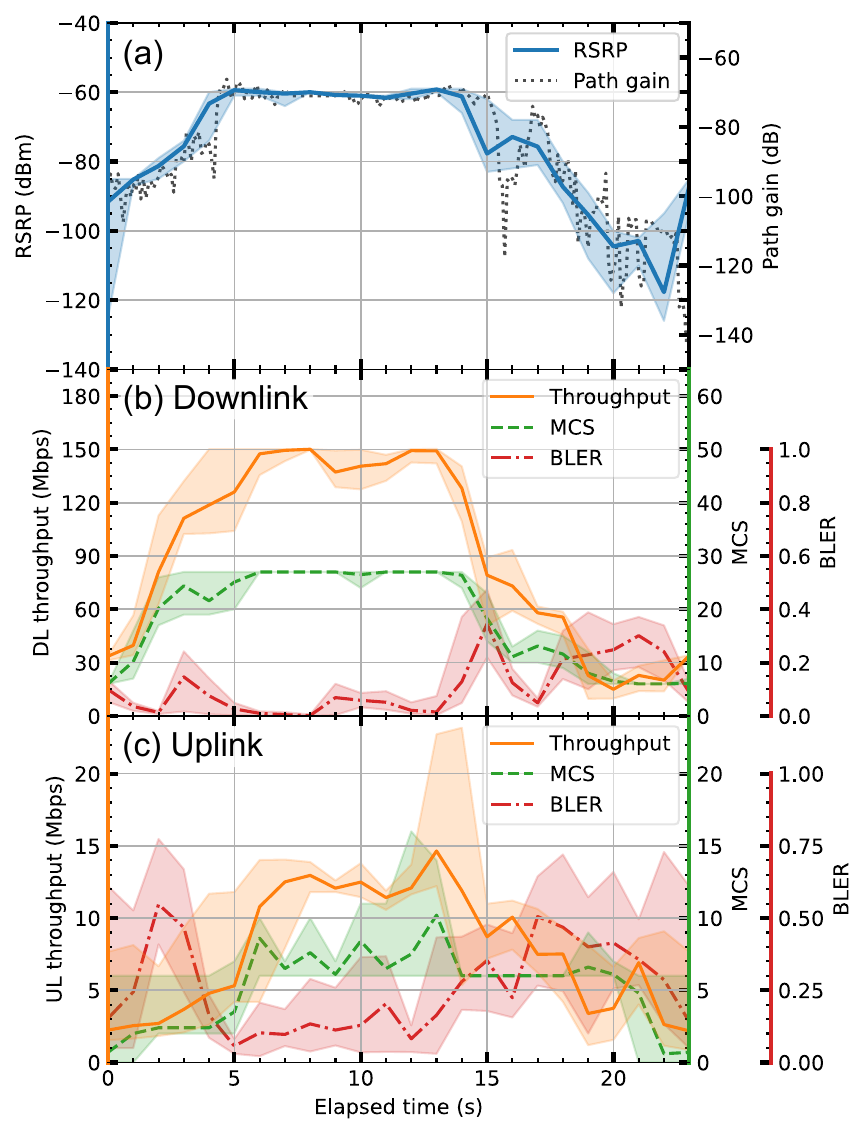}}
\caption{(a) RSRP with path gain obtained from RT simulation as a reference, (b) DL KPIs, and (c) UL KPIs obtained from OAI for Shibuya scenario.\label{fig:shibuyaOAI}}
\end{figure}
Figs.~\ref{fig:shinnakanoOAI} and \ref{fig:shibuyaOAI} summarize the time evolution of various KPIs output by the OAI during the execution of the Shin-nakano and Shibuya scenarios, respectively. The bands shown in lighter shades of the same color as each plot represent the upper and lower bounds observed when the corresponding scenario was executed ten times, indicating the range of error margins.

For the CIR data generated in the previous section, the value of $s^2$ was configured such that ($5-$[the maximum path gain within the scenario])~dB was achieved in (\ref{eq:yk}) to maintain a stable connection without interruptions in the KPI measurements throughout the scenario. Additionally, because noise power was not the primary focus of this study, $\sigma^2$ was set to -100~dBm. Therefore, it should be noted that the absolute values of the observed RSRP in this study are not necessarily realistic, and emphasis is placed on relative changes in RSRP throughout the scenario.

First, the variations in the RSRP obtained from the OAI and path gain from the Sionna RT, as shown in Figs.~\ref{fig:shinnakanoOAI}(a) and \ref{fig:shibuyaOAI}(a), exhibit a generally consistent trend. For example, in both scenarios, a plateau in the RSRP was observed during the red traffic light periods for both the DL and UL. However, the short-term fluctuations of approximately 20~dB and the long-term variations in the 18-44 second interval in the path gain of the Shin-nakano scenario are not fully reflected in the observed RSRP. This discrepancy is likely due to the KPI averaging process performed by the OAI gNB and the automatic gain control (AGC) function that may be implemented in COTS UE. As a result, particularly in the Shin-nakano scenario, a level adjustment of approximately 20-30~dB might have occurred within the aforementioned time interval.

Furthermore, the lower sections of these figures show that the MCS, BLER, and throughput fluctuate in response to time variations in the RSRP. In this section, we quantitatively examine the throughput. The maximum bitrate $R_\text{b}$ [Mbps] of 5G NR can be estimated using the following formula, as specified in 3GPP TS 38.306~\cite{bitrate}.
\begin{align}
    R_\text{b} &= 10^{-6}\ Q_{m} R_\text{max} \frac{12\ N^{BW}_\text{PRB}}{T_\text{symb}}\left(1-OH\right) \\
    &=
    \left\{
    \begin{array}{ll}
    30.62976\ \eta & \text{(for DL)} \\
    32.76672\ \eta & \text{(for UL)}. \label{eq:bitrate}
    \end{array}
    \right.
\end{align}
Note that the notation is simplified by assuming the number of MIMO layers as 1, the number of aggregated component carriers as 1, and the scaling factor to be 1. Here, the modulation order $Q_m$, coding rate $R_\text{max}$, and their product, the spectral efficiency $\eta$ are uniquely determined by the MCS index in TS 38.214 Table 5.1.3.1-2\cite{MCS}. $N^\text{BW}_\text{PRB}$ represents the number of physical resource blocks (PRBs) within $BW$, which is determined by the frequency range and the numerology. $T_\text{symb}$ is the average OFDM symbol duration in a subframe, and $OH$ represents the control channel overhead. Some variables are determined by the parameters in Table \ref{table:param} and remain unchanged throughout the scenario as $N^{BW}_\text{PRB} = 106$, $T_\text{symb} = 3.57$ [\textmu s], and $OH$ is 0.14 for DL and 0.08 for UL. By substituting these specific values, the resulting equation is given by (\ref{eq:bitrate}). The effective throughput for DL (UL) $T^\text{eff}_\text{DL(UL)}$ is defined as the actual received data rate, which is given in the case of TDD mode by
\begin{equation}
    T^\text{eff}_\text{DL(UL)} = (1 - BLER)\ \alpha_\text{DL(UL)} R_\text{b},
\end{equation}
where $BLER$ represents the block error rate defined by the ratio of the number of erroneous blocks to the total number of transmitted blocks, and $\alpha_\text{DL(UL)}$ indicates the time occupancy ratio of DL (UL) determined by the TDD slot pattern and special slot format. In the case of the TDD pattern shown in Table \ref{table:param}, the DL ratio is given by $\alpha_\text{DL}=0.6857$, while the UL ratio is given by $\alpha_\text{UL}=0.2571$, and thus, $T^\text{eff}_\text{DL(UL)}$ can be expressed as follows:
\begin{align}
    T^\text{eff}_\text{DL} &= 21.003264\ (1 - BLER)\ \eta \label{eq:DLThr}\\
    T^\text{eff}_\text{UL} &= 8.425728\ (1 - BLER)\ \eta. \label{eq:ULThr}
\end{align}
Therefore, it can be concluded that the effective throughput can be calculated based on the spectral efficiency $\eta$ corresponding to the MCS index and $BLER$. Accordingly, we focus on the variations in the MCS and $BLER$ depending on the mobility scenario. In the two scenarios analyzed, the maximum observed DL throughput was 150.0778~Mbps. At this point, the MCS $=27$ and $BLER=0.001656$. Using (\ref{eq:DLThr}), the calculated effective throughput is 155.2989~Mbps, showing a close match between measurement and calculation. On the other hand, the maximum observed UL throughput was 14.64214~Mbps. At this point, the MCS $=10$, and the $BLER=0.163917$. Using (\ref{eq:ULThr}), the calculated effective throughput is 18.106756~Mbps. The discrepancy in UL throughput compared with DL throughput can be attributed to the presence of sounding reference signals (SRSs), which occupy a portion of the PRBs in UL transmission. Therefore, the KPIs measured by KPM xApps can be considered consistent with the theoretical expectations. The measured throughputs showed a maximum when the UE was closest to the gNB, the BLER exhibited an inverse trend to the throughput, and the MCS changed in a near-proportional relationship to the throughput, indicating reasonable results. When comparing Figs.~\ref{fig:shinnakanoOAI}(b) and (c), examining the difference between the DL and UL, it becomes apparent that UL variations are not as pronounced as those in the DL. The BLER is significantly higher, and to maintain UL communication quality, the MCS is automatically set lower to limit the modulation order, resulting in lower UL throughput. The potential causes of errors include (i) modulation and demodulation methods, (ii) modulation symbol period, (iii) characteristics of bandwidth limitation filters, (iv) channel characteristics, and (v) timing offsets due to cycle slips. These errors originate from both system factors and radio propagation issues. If we focus solely on the propagation aspects, the relevant factors are (iv) and (v). Because this study does not consider the velocity vector of the vehicle, the possibility of cycle slips caused by fast fading is eliminated. Consequently, inter-symbol interference (ISI) errors caused by multipath propagation with varying arrival times remain a primary concern. The Cyclic Prefix (CP) in CP-OFDM, used in 5G NR, is a technique designed to mitigate ISI. In our communication system parameters, the long CP assigned to the first OFDM symbol contains 132 samples corresponding to 2.86 \textmu s, while subsequent short CPs contain 106 samples corresponding to 2.30 \textmu s. If significant complex amplitudes are found in taps exceeding 132 samples within the 146 CIR taps used in this study, the throughput could be affected. In our environment, the real-time processing limit was $L_\text{OAI}=28$ taps, and no delay tap index exceeding 2.30 \textmu s was observed. Thus, it is concluded that the ISI due to multipath propagation is not an issue in these scenarios, and the error observed in the UL is considered to originate from the system. As shown in Table \ref{table:baseline}, the UL BLER was already close to 10\% during the baseline measurement, which may be related to the implementation of received signal processing on the gNB.

To assess the resilience of the CP-OFDM waveform used by the OAI to the largest Doppler shift expected in our mobility scenario, we rely on the well-known set of conditions that must be satisfied for reliable OFDM transmission \cite{Karasawa}:
\begin{equation*}
    \sigma_{\tau}\ll T_\text{GI}\ll T_\text{OFDM}\ll T_\text{f}\left(=\frac{1}{f_\text{D}}=\frac{c}{fv}\right),
\end{equation*}
where $\sigma_{\tau}$ is the RMS delay spread, $T_\text{GI}$ is the guard-interval (CP) length, $T_\text{OFDM}$ is the effective OFDM symbol duration (excluding the CP), $T_\text{f}$ is the fading (phase variation) period, $f_\text{D}$ is the maximum Doppler frequency, $c$ is the speed of light, and $v$ is the maximum UE speed. 
First, as discussed above, no multipath component with a delay exceeding the short CP length $T_\text{GI}=2.30$ [\textmu s] was observed; therefore, $\sigma_{\tau}\ll T_\text{GI}$ holds, and the ISI is avoided. Second, the ratio $T_\text{GI}/T_\text{OFDM}$ (2.30 \textmu s / 33.3 \textmu s) follows the 5G NR standard, meaning $T_\text{GI}\ll T_\text{OFDM}$ and high spectral efficiency are maintained. Third, with the maximum vehicle speed $v=11.78$ [m/s], giving
\begin{equation*}
    T_\text{f}=\frac{c}{fv}=\frac{2.998\times 10^8}{4.01916\times10^9\times11.78}=6.332 \text{ [ms]},
\end{equation*}
which is far longer than $T_\text{OFDM}$; hence the phase coherence condition $T_\text{OFDM}\ll T_\text{f}$ is also satisfied. Because all three inequalities are true, the Doppler shift $f_\text{D}=158$ [Hz] has a negligible effect on the link quality for CP-OFDM transmission under the parameters set in this study. At higher carrier frequencies (e.g., FR2) and high-speed mobility scenarios (e.g., trains), these conditions become far more stringent, and severe ISI or fast fading is expected. Because the shortest fading period is $T_\text{f}=6.332$ [ms], the CIR snapshot interval should, in principle, be no larger than $T_\text{f}$. An efficient approach is to compute snapshots every approximately 20~ms (not exceeding several times $T_\text{f}$) and then use Sionna RT’s \texttt{apply\_doppler()} function to interpolate the Doppler-affected CIR at a finer time resolution. Capturing the $6.332$~ms periodicity only requires interpolation at $\sim1$~ms steps; therefore, the CIR data can be pre-generated offline. The remaining issue is to convolve these time-varying CIRs with the baseband signal in real time. As the effective CIR interval shortens, the data loading overhead increases during real-time processing, requiring additional acceleration techniques.

\subsection{FUTURE PROSPECTS}
When discussing future intelligent transportation systems, it is necessary to consider methods for appropriately evaluating mobility scenarios in which the statistical properties of the channels change rapidly over time in dynamic radio environments. Although ray tracing achieves high-resolution results in static channel modeling, it becomes significantly more complex in dynamic scenarios involving high-speed moving vehicles, where the wide-sense stationary (WSS) assumption is violated. Generally, frequency- and time-selective fading must be considered for mobile entities in multipath environments. However, realistically incorporating the effects of both types of fading into baseband signal processing and executing real-time E2E wireless communication are extremely challenging because of the computational load. To address this challenge, we aim to explore real-time processing, including support for Doppler effects, using acceleration technologies such as FPGAs and GPUs in future studies. 

From a scaling perspective covering multiple gNB/UE nodes and MIMO emulation, we believe that the most promising approach is to offload the convolution operation to an FPGA board inserted into the cabled link between the gNB and UE. In this setup, the FPGA acts as a channel emulator that can process up to [the number of (\#) RF ports on the board / 2] simultaneous channels (one TX/RX pair per channel). Accordingly, scalability is available as long as:
\begin{equation*}
[\text{\#RF ports}]/2 \geq [\text{\#gNBs}]\times[\text{\#UEs}]\times[\text{\#MIMO layers}].
\end{equation*}
Because convolution is well suited to the low-latency parallel architecture of an FPGA, we consider hardware offloading as an effective means to achieve scalable emulation.

Beamforming for higher frequency bands, such as FR2, also necessitates tracking moving targets, which may result in frequent handovers. Because a ray-tracing simulator can account for both the beam pattern of the gNB and the antenna pattern of the UE, it is possible to create FR2 mobility scenarios in which the beam dynamically tracks a moving UE. Once such a scenario is generated, the OWDT enables the development of a solution for managing handovers throughout the entire system.

To achieve these objectives, the following extensions are required: integration with higher-precision 3D models and mobility scenarios, scalability in terms of the number of gNBs and UEs, utilization of acceleration technologies such as GPUs and FPGAs to support increased computational complexity, and validation through a comparison with experimental data to establish a truly representative digital twin.

\section{CONCLUSION}\label{conclusion}
This study proposed the OWDT, an E2E 5G mobility emulation framework that can be deployed as an O-RAN ecosystem. By combining the OAI for protocol stack emulation and Sionna RT for ray-tracing-based RF modeling, the OWDT enables a high-fidelity performance evaluation in vehicular-mobility scenarios. To achieve real-time emulation, we implemented a convolution method that applies pre-computed time-evolving CIRs to baseband IQ signals. This method efficiently reproduces multipath fading effects while ensuring real-time processing within the constraints of COTS hardware. The framework leverages near-RT RIC via FlexRIC to monitor KPIs, such as RSRP, MCS, BLER, and throughput, providing valuable insights into the behavior of wireless communication systems. Experimental validation demonstrated that the OWDT effectively replicates real-world channel conditions, supporting efficient network performance analysis and optimization with minimal physical deployments. This approach significantly reduces costs and accelerates the development of wireless communication systems. 
A key feature of this study is that all workflows and system architectures were built entirely using OSS and open data. The OWDT architecture we constructed serves as a reference architecture, and each component can be substituted with any proprietary product or alternative OSS implementation.

\section*{ACKNOWLEDGMENT}
T. Iye thanks Luis Pereira and Paulo Marques from Allbesmart LDA for the fruitful discussion on the OAI implementation. 
The authors used generative AI tools, such as ChatGPT, Grammarly, and Paperpal, to improve the grammar of the manuscript. After using these tools, they carefully reviewed and edited the content and took full responsibility for the final published manuscript.

\bibliographystyle{IEEEtran}
\bibliography{reference}

\begin{IEEEbiography}[{\includegraphics[width=1in,height=1.25in,clip,keepaspectratio]{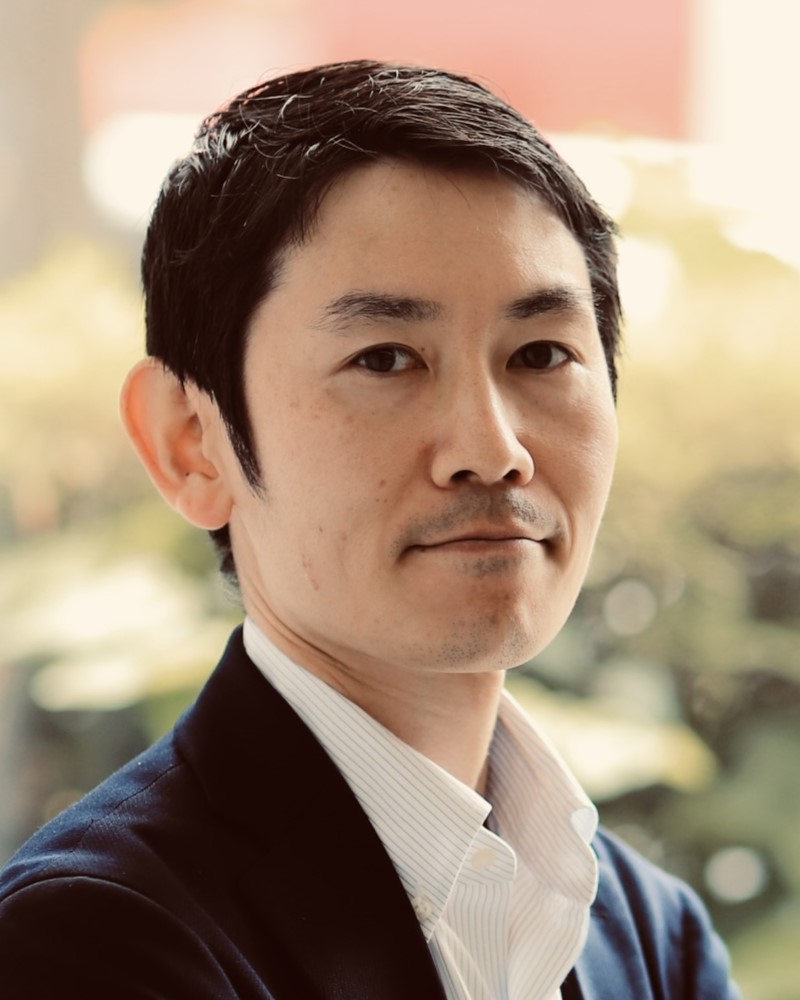}}]{TETSUYA IYE } received the B.S. degree in physics from Waseda University, Tokyo, Japan, in 2009 and the M.S. and Ph.D. degrees in physics from Kyoto University, Kyoto, Japan, in 2011 and 2014, respectively. From 2013 to 2014, he was a Research Fellow of the Japan Society for the Promotion of Science. He joined Kozo Keikaku Engineering Inc., Tokyo, Japan, as a system engineer in 2014. After gaining experience in developing CAD-related systems, vibration analysis, and IoT system development, he has been engaged in the research and development of next-generation mobile communication systems since 2019. His research interests include software-defined radio, antennas and propagation, signal processing, and AI/ML. He is a member of the IEICE.
\end{IEEEbiography}

\begin{IEEEbiography}[{\includegraphics[width=1in,height=1.25in,clip,keepaspectratio]{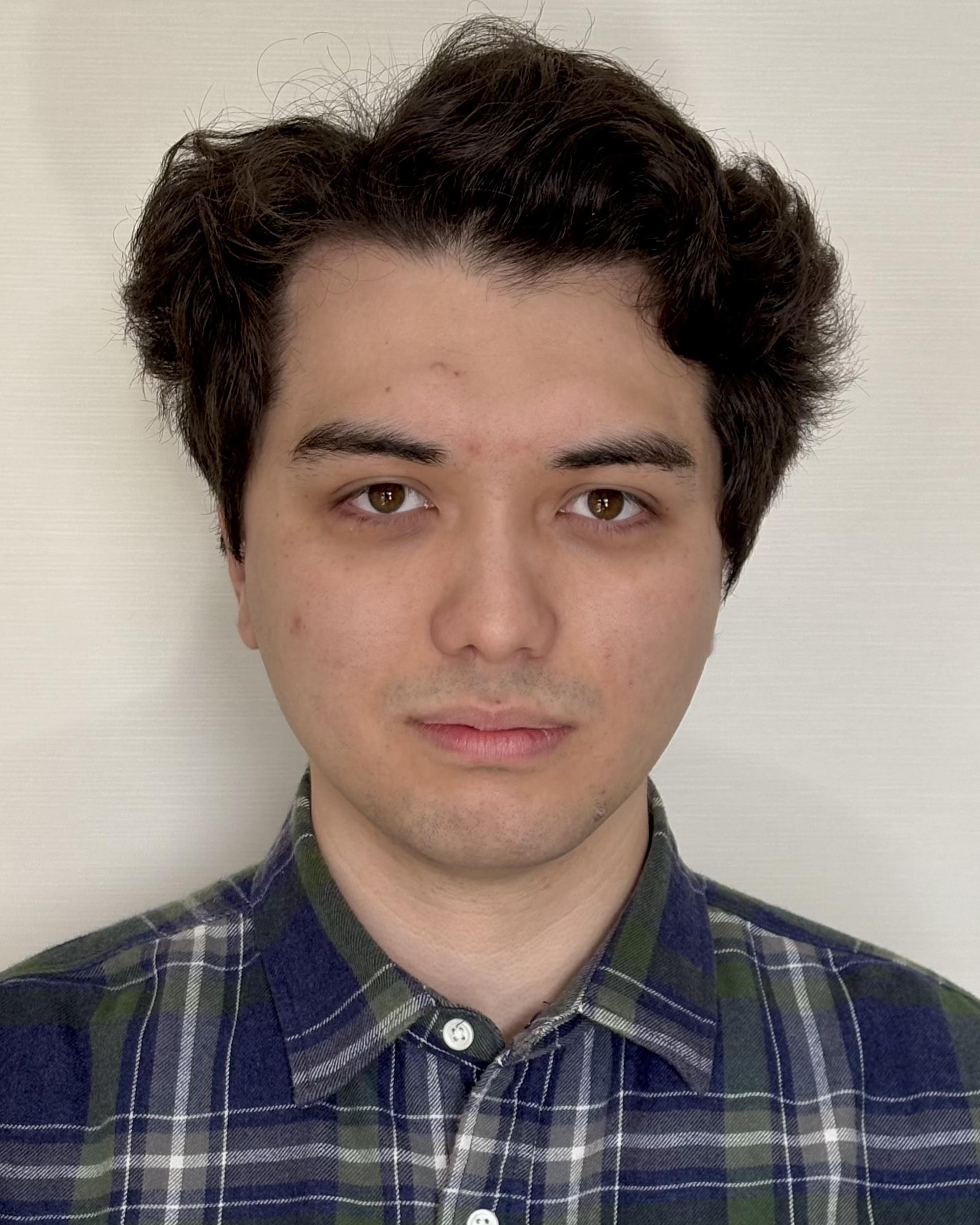}}]{MASAYA SAKAMOTO } received the B.S. in physics from Tokyo Metropolitan University, Tokyo, Japan, in 2024. In the same year, he began collaborating with the Communication Engineering Department at Kozo Keikaku Engineering Inc., Tokyo, Japan, as an independent research engineer specializing in data analysis and algorithm design. His research interests include deep learning and geometric properties of hidden layers.
\end{IEEEbiography}

\begin{IEEEbiography}[{\includegraphics[width=1in,height=1.25in,clip,keepaspectratio]{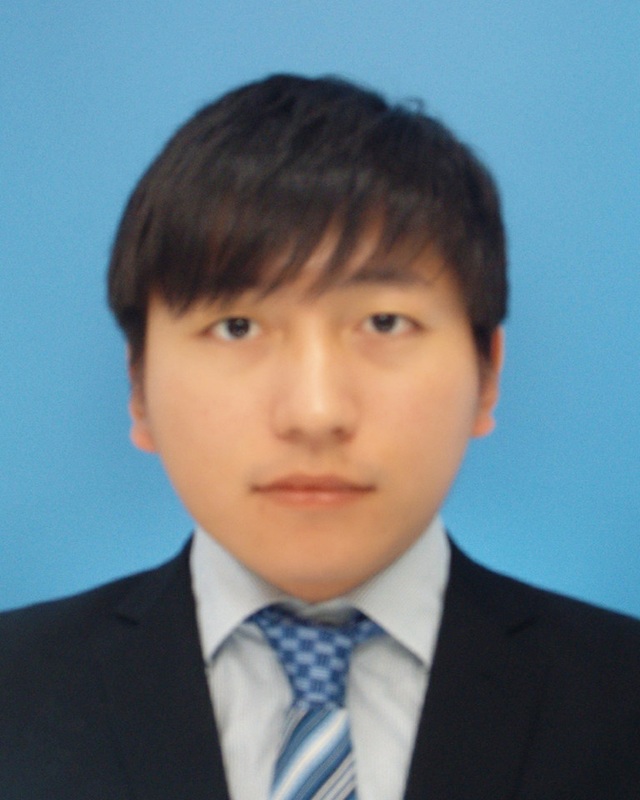}}]{SHOHEI TAKAYA } received the B.E. degree in electrical and electronic engineering from Tokyo Denki University, Japan, in 2016. From 2016 to 2020, he worked at Miraxia Edge Technology Corporation. Since 2021, he has been working at Kozo Keikaku Engineering, Inc. He has been researching and developing 5G systems and radio propagation emulation using software-defined radios. He is a member of the IEICE.
\end{IEEEbiography}

\begin{IEEEbiography}[{\includegraphics[width=1in,height=1.25in,clip,keepaspectratio]{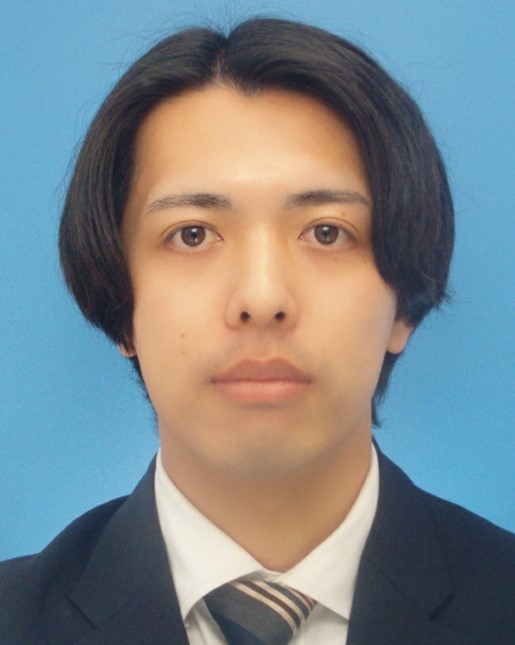}}]{EISAKU SATO } received the B.E. and M.E. degrees from Toyohashi University of Technology, Japan, in 2021 and 2023, respectively. In 2023, he joined Kozo Keikaku Engineering Inc., Japan, as a wireless communication engineer. His current research and development interests include system-level simulation and wireless signal processing. He is a member of the IEICE.
\end{IEEEbiography}

\begin{IEEEbiography}[{\includegraphics[width=1in,height=1.25in,clip,keepaspectratio]{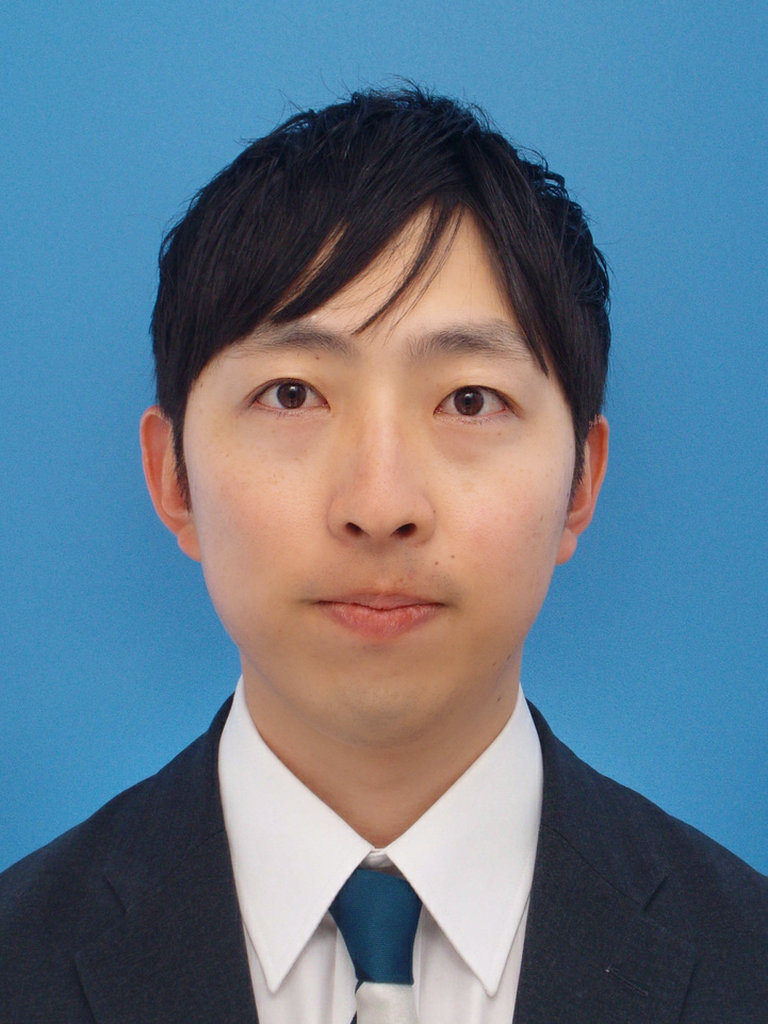}}]{YUKI SUSUKIDA } received the B.S. and M.S. degrees in Seismology from Hokkaido University, Japan, in 2020. Since 2020, he has been working at Kozo Keikaku Engineering Inc., where he focuses on the research and development of wireless access systems using software-defined radio technologies and OpenAirInterface, an open-source software platform for 5G communications systems. His research interests include 5G V2X, 5G NTN, and Integrated Sensing and Communication (ISAC).
\end{IEEEbiography}

\begin{IEEEbiography}[{\includegraphics[width=1in,height=1.25in,clip,keepaspectratio]{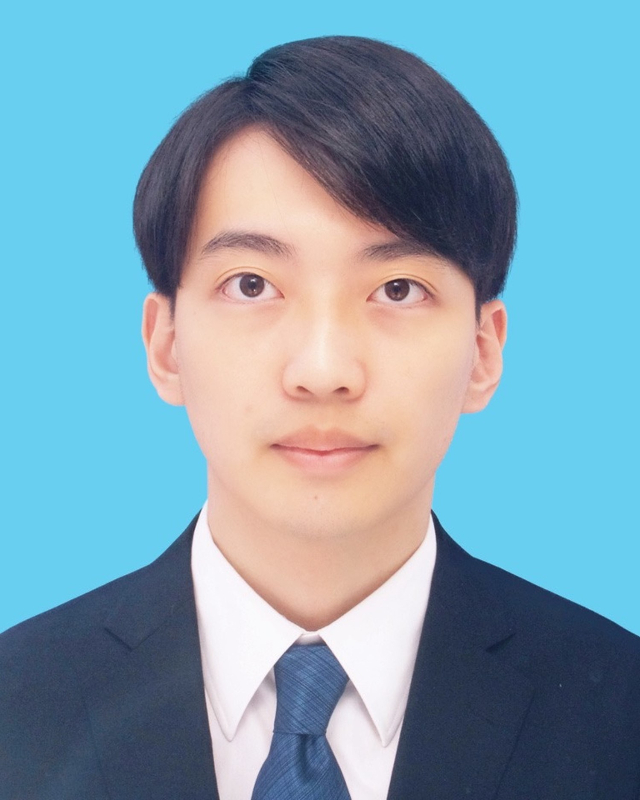}}]{YU NAGAOKA } received the B.E. degree in Electronics and Computer Systems from Takushoku University, Japan, in 2024. Since 2024, he has been pursuing an M.E. degree in Mechanical and Electronic Systems Engineering at the Graduate School of Engineering, Takushoku University. His current research interests include wireless communication systems and their protocols. He is a member of the IEICE.
\end{IEEEbiography}

\begin{IEEEbiography}[{\includegraphics[width=1in,height=1.25in,clip,keepaspectratio]{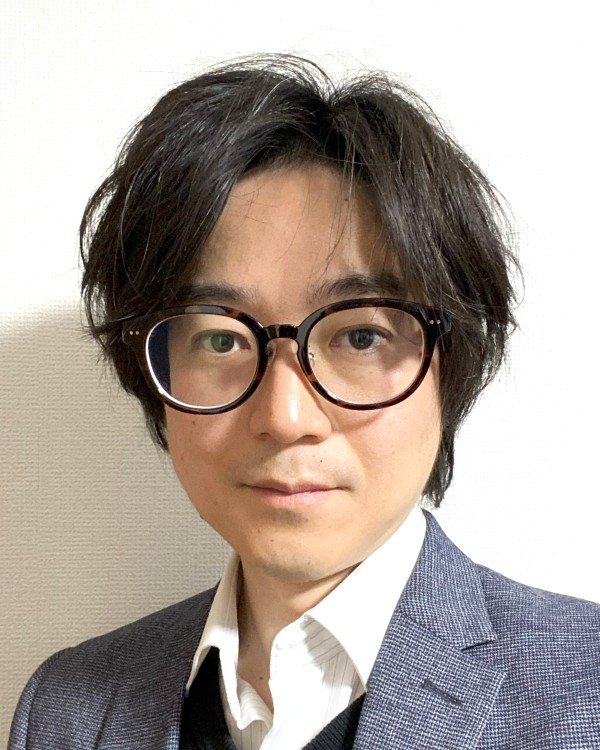}}]{KAZUKI MARUTA } (Senior Member, IEEE) received the B.E., M.E., and Ph.D. degrees in engineering from Kyushu University, Japan in 2006, 2008, and 2016, respectively. 
From 2008 to 2017, he was with NTT Access Network Service Systems Laboratories.
From 2017 to 2020, he was an Assistant Professor at the Graduate School of Engineering, Chiba University. 
From 2020 to 2022, he was a specially appointed Associate Professor at the Academy for Super Smart Society, Tokyo Institute of Technology. 
He is currently an Associate Professor at the Department of Electrical Engineering, Tokyo University of Science. 
His research interests include MIMO, massive MIMO, adaptive array signal processing, channel estimation, moving networks, visible light communication, and underwater acoustic communication. 
He is a senior member of the IEEE and the IEICE. 
He serves as a Poster Co-Chair for IEEE CCNC from 2023 to 2025. 
He received the IEICE Young Researcher's Award in 2012, the IEICE Radio Communication Systems Active Researcher Award in 2014, 
APMC2014 Prize, the IEICE RCS Outstanding Researcher Award in 2018, the IEEE ICCE Excellent Paper Award in 2021, and the ICAIIC 2024 Excellent Paper Award. 
He was a co-recipient of the IEICE Best Paper Award in 2018, the APCC 2019 Best Paper Award, the ICCE-Asia 2022 Best Paper Award, and the ICUFN 2025 Best Paper Award.
\end{IEEEbiography}

\begin{IEEEbiography}[{\includegraphics[width=1in,height=1.25in,clip,keepaspectratio]{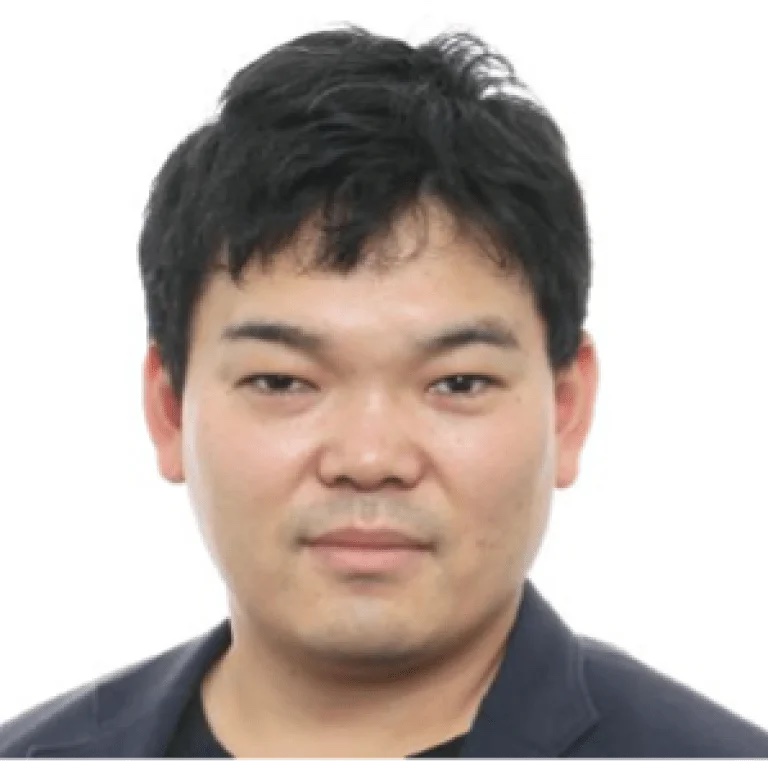}}]{JIN NAKAZATO } (Member, IEEE) is currently an Assistant Professor at Tokyo University of Science. He also works as a technical advisor with Visban Corporation. He received B.E. and M.E. degrees from the University of Electro-Communications, Japan, in 2014 and 2016, respectively. He received a Ph.D. degree from the Tokyo Institute of Technology, Japan, in 2022. From 2016 to 2020, he was with Fujitsu Limited. From 2020 to 2022, he was with Rakuten Mobile, Inc. From 2022 to 2024, he was with specially appointed Assistant Professor at the University of Tokyo. His research interests include Multi-access Edge Computing, NFV/SDN Orchestrator, V2X, Open RAN, UAV networks, and virtualization RAN. He is a Member of IEICE and IEEE. He received the Best Paper Award at the International Conference on Ubiquitous and Future Networks (ICUFN) in 2019 and 2024. He also received the Best Paper Award at the INFOCOM 2024, ICAIIC Excellent Paper Award, and ACM ICEA Best Paper Award. He serves as a peer-reviewed open-access letter journal covering the field of communication, International Journal of Computers Applications Associate Editor, and Computer Networks Software and Datasets Editors.
\end{IEEEbiography}
\EOD
\end{document}